\documentclass[journal]{IEEEtran}

\pdfoutput=1

\usepackage{amsmath,amssymb,amsfonts}
\usepackage{algorithmic}
\usepackage{textcomp}
\usepackage{xcolor}
\usepackage{bm}
\usepackage{float}
\usepackage{booktabs,multirow}
\usepackage{cite}
\usepackage{acronym}
\usepackage{tikz}
\usepackage{gensymb}
\usepackage[group-separator={,}]{siunitx}

\usepackage{subfig}  
\usepackage{mathtools}

\usepackage[bookmarks=false,colorlinks=true,citecolor=mygreen!60!black,linkcolor=myblue]{hyperref}
\usepackage[capitalize]{cleveref}  

\usepackage{tikz}
\newcommand{\tikzcircle}[2][black,fill=black]{\tikz[baseline=-0.5ex]\draw[#1,radius=#2] (0,0) circle ;}%

\definecolor{mygreen}{rgb}{0.6000,0.7529,0}%
\definecolor{myblue}{rgb}{0 0.3529 0.6627}%
\definecolor{myyellow}{rgb}{1.0000,0.8784,0.3608}%
\definecolor{myred}{rgb}{0.9020,0,0.1020}%
\definecolor{myturkis}{rgb}{0,0.6157,0.5059}%

\acrodef{GRF}{ground reaction forces}
\acrodef{3D}{three-dimensional}
\acrodef{EM}{electromagnetic}
\acrodef{LOS}{line-of-sight}

\newcommand{\Hyp}{\mathcal{H}}

\begin{document}
\title{Doppler Radar for the Extraction of \\Biomechanical Parameters in Gait Analysis}

\author{Ann-Kathrin~Seifert,~\IEEEmembership{Student Member,~IEEE,}
           Martin~Grimmer,
           and Abdelhak~M.~Zoubir,~\IEEEmembership{Fellow,~IEEE}
\thanks{\textbf{\textcopyright~2020 IEEE. Personal use of this material is permitted. Permission from IEEE must be obtained for all other uses, in any current or future media, including reprinting/republishing this material for advertising or promotional purposes, creating new collective works, for resale or redistribution to servers or lists, or reuse of any copyrighted component of this work in other works..}

A.-K.~Seifert and A.~M.~Zoubir are with the Signal Processing Group at Technische Universit\"at Darmstadt, Darmstadt, Germany (correspondence e-mail: seifert@spg.tu-darmstadt.de).

M.~Grimmer is with the Institute of Sports Science at Technische Universit\"at Darmstadt, Darmstadt, Germany.

The work of M.~Grimmer was funded by the German Science Foundation (DFG) under the grant number GR 4689/3-1.
}
}

\maketitle

\begin{abstract}
The applicability of Doppler radar for gait analysis is investigated by quantitatively comparing the measured biomechanical parameters to those obtained using motion capturing and ground reaction forces. 
Nineteen individuals walked on a treadmill at two different speeds, where a radar system was positioned in front of or behind the subject. The right knee angle was confined by an adjustable orthosis in five different degrees.
Eleven gait parameters are extracted from radar micro-Doppler signatures. Here, new methods for obtaining the velocities of individual lower limb joints are proposed. Further, a new method to extract individual leg flight times from radar data is introduced. Based on radar data, five spatiotemporal parameters related to rhythm and pace could reliably be extracted. Further, for most of the considered conditions, three kinematic parameters could accurately be measured. The radar-based stance and flight time measurements rely on the correct detection of the time instant of maximal knee velocity during the gait cycle. This time instant is reliably detected when the radar has a back view, but is underestimated when the radar is positioned in front of the subject. The results validate the applicability of Doppler radar to accurately measure a variety of medically relevant gait parameters. Radar has the potential to unobtrusively diagnose changes in gait, e.g., to design training in prevention and rehabilitation. As contact-less and privacy-preserving sensor, radar presents a viable technology to supplement existing gait analysis tools for long-term in-home examinations.
\end{abstract} 

\begin{IEEEkeywords}
Doppler radar, motion capture, gait analysis, ambient assisted living
\end{IEEEkeywords}
  
\IEEEpeerreviewmaketitle

\section{Introduction}

\IEEEPARstart{G}{ait} analysis encompasses the measurement and assessment of quantities that characterize human locomotion. Such analysis is not only important for biomechanical research, sports medicine and athletic training, but it can also reveal information on the general state of individual health. For a variety of diseases, clinical research has shown that gait analysis can facilitate diagnosis and enhance patient treatment. Pathologies affecting the gait include neurodevelopmental disorders (e.g.~cerebral palsy and down syndrome) as well as neurodegenerative disorders (e.g.~Parkinsons's disease and Alzheimer's) \cite{Jar18,Gho17,Che16,Wah15}. Further, gait analysis is employed in orthopedics, prosthetics and rehabilitation (e.g.~post stroke) as well as geriatrics (e.g.~fall risk assessment) \cite{Jar18,Che16,Gri19a,Bar06}.

A great variety of sensors have been proposed for analyzing the gait. Here, we focus on systems that are capable of providing continuous, unobtrusive, and pervasive locomotion monitoring in home, e.g., to support aging in place and enable independence in assisted living \cite{Deb16}. These systems can broadly be categorized into wearable (direct) (for reviews of wearable sensors see e.g.~\cite{Jar18,Che16}) and non-wearable (indirect) solutions (for an overview see e.g.~\cite{Mur14}). However, wearable sensors require precise placement, necessitate the cooperation of the users and can be uncomfortable to wear, which might also affect the gait. For these reasons non-wearable technologies are desirable. Marker-less motion capture systems such as camera-based systems operating at visible light \cite{Cai17,Wan13} and/or infrared light \cite{Hag10}, including the use of Microsoft Kinect (see e.g.~\cite{Lat19,Gho17,Xu15,Ero15,Wah15,Pfi14,Sto13}), have widely been investigated for gait analysis (for a review on vision-based motion analysis see e.g.~\cite{Col18}). However, gait monitoring systems based on video cameras can be affected by poor lighting conditions and clothing, and are generally not privacy-preserving and thus may not be accepted as an in-home modality. Depth-based systems based on infrared light often require the entire body to be in the field of view and typically model the human posture by locating a finite number of points on the skeleton, which might be insufficient to capture detailed gait motions. In order to account for these drawbacks, we propose the use of radar as a marker-less remote sensor for gait analysis. 

Radar has widely been used in automotive scenarios \cite{Eng17,Pat17}, for human detection behind walls with application to search and rescue operations \cite{Deb12,Lei15}, as well as for health monitoring, e.g., vital sign monitoring \cite{Li13}. More recently, radar technology has gained much attention in the field human activity recognition (HAR) for activities of daily living (ADL) \cite{Lek19,Ami17}. Since it is safe, contact-less, privacy-preserving, and does not require any instrumentation on the user, it is a viable sensor for continuous and unobtrusive in-home gait analysis, which could be used in conjunction to other wearable and remote sensors. In order to use radar systems as medical devices, i.e., for sensing small alternations in gait patterns that can be indicative for declining functional health, these changes need to be quantified. Opposed to automatic data-driven feature learning using e.g.~neural networks (see e.g.~\cite{Van18,Sey18,Che18,Kim16}), we extract spatiotemporal and kinematic gait parameters that are valuable to clinicians for assessing the gait (see e.g.~\cite{Sev17,Ser19,Wah15} as well as references in \cite{Jar18}). Relevant prior research on radar-based gait analysis includes the work of Wang \textit{et al.} \cite{Wan14}, who employed pulse-Doppler radar to estimate the step time as well as the walking speed of thirteen volunteers. Based on radar micro-Doppler signatures of $74$ seniors, Saho \textit{et al.} \cite{Sah19} showed that not only the walking speed, but also leg and foot velocities are indicative for lower cognitive functions in the elderly. 

In this paper, we qualitatively and quantitatively compare radar-based gait parameters to simultaneously recorded marker-based motion capture data utilizing ground reaction forces from an instrumented treadmill. Marker-based motion capture data have previously been utilized for interpreting and simulating micro-Doppler gait signatures \cite{Ram17,Che11}. For example, in automotive scenarios, Held \textit{et al.} \cite{Hel18} used motion capture data for analyzing micro-Doppler signatures of human gait for enhanced pedestrian detection and early movement prediction. Karabacak \textit{et al.} \cite{Kar15} utilized motion capture data and the frequently used global human motion model by Boulic, Magnenat-Thalman and Thalman \cite{Bou90} to simulate micro-Doppler gait signatures and validated their use for classification. Abdulatif \textit{et al.} \cite{Abd17} used motion-capture data to develop an improved motion model to track individual body components based on radar measurements. 

However, most of these works merely qualitatively discuss the consistency between radar measurements and (simultaneously) recorded motion capture data. In this paper, we utilize motion capture data to validate existing and new methods for extracting a variety of gait parameters from experimental radar data, including spatiotemporal and kinematic parameters. A new method to extract the swing phase (flight time) of individual legs from the radar data is introduced. Additionally, we propose new methods for obtaining sagittal velocities of individual lower limb joints, i.e., toe, ankle and knee velocities. Based on experimental data of nineteen able-bodied volunteers walking on a treadmill, we investigate the capabilities of radar to capture gait parameters in unimpaired gait and for gait abnormalities that were introduced by an adjustable orthosis, such that one of the knees could not be fully bent. Further, we present results for two different walking speeds ($0.7$\,m/s and $1.1$\,m/s) and discuss the differences in positioning the radar in front of or behind the test subject. 

\section{System description and data collection}\label{sec:setup}

\subsection{Experimental setup}
\begin{figure}[!t]
    \centering
    \includegraphics[width=\columnwidth]{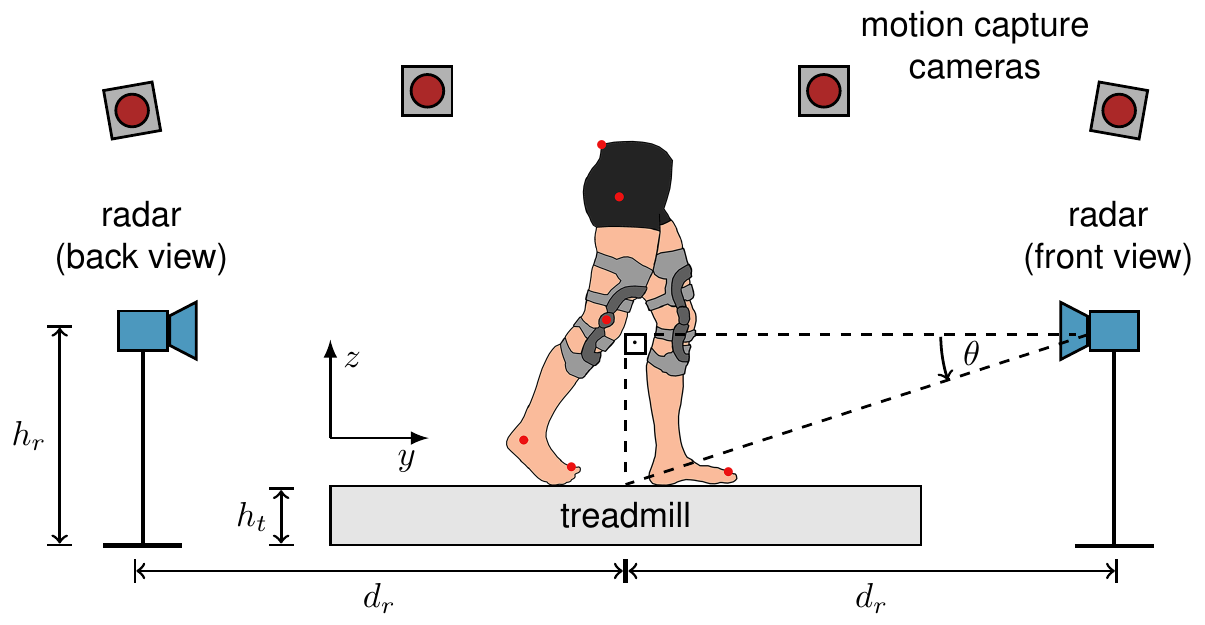}
    \caption{Schematic of the experimental setup with an instrumented treadmill that measures \acp{GRF}, a motion-capture system with $12$ infrared cameras, and two radar systems (whereof only one was active at a time). The red dots indicate the positions of the reflective markers for motion capturing.}
    \label{fig:exp_setup}
\end{figure}

\cref{fig:exp_setup} schematically shows the experimental setup.The radar data were collected using a continuous-wave radar (SDR-KIT 2400AD, Ancortek, Fairfax, VA, USA) with a transmitting frequency of $24$\,GHz \cite{Anc}. The horn antennas, one transmitting and one receiving, were positioned at approximately knee height, i.e., $h_r - h_t = 0.58$\,m above the treadmill surface, and $d_r = 1.75$\,m in front of or behind the center of the treadmill. At the center of the treadmill, the radar's 3dB beam width covers the vertical range from approximately $12$ to $104$\,cm above the treadmill surface. Further, the radar system was positioned such that the test subjects walked in a \ang{0}~angle to the radar's \ac{LOS}. Only one radar system was active at a time.

\begin{figure}[!t]
    \centering
    \includegraphics[width=0.49\columnwidth]{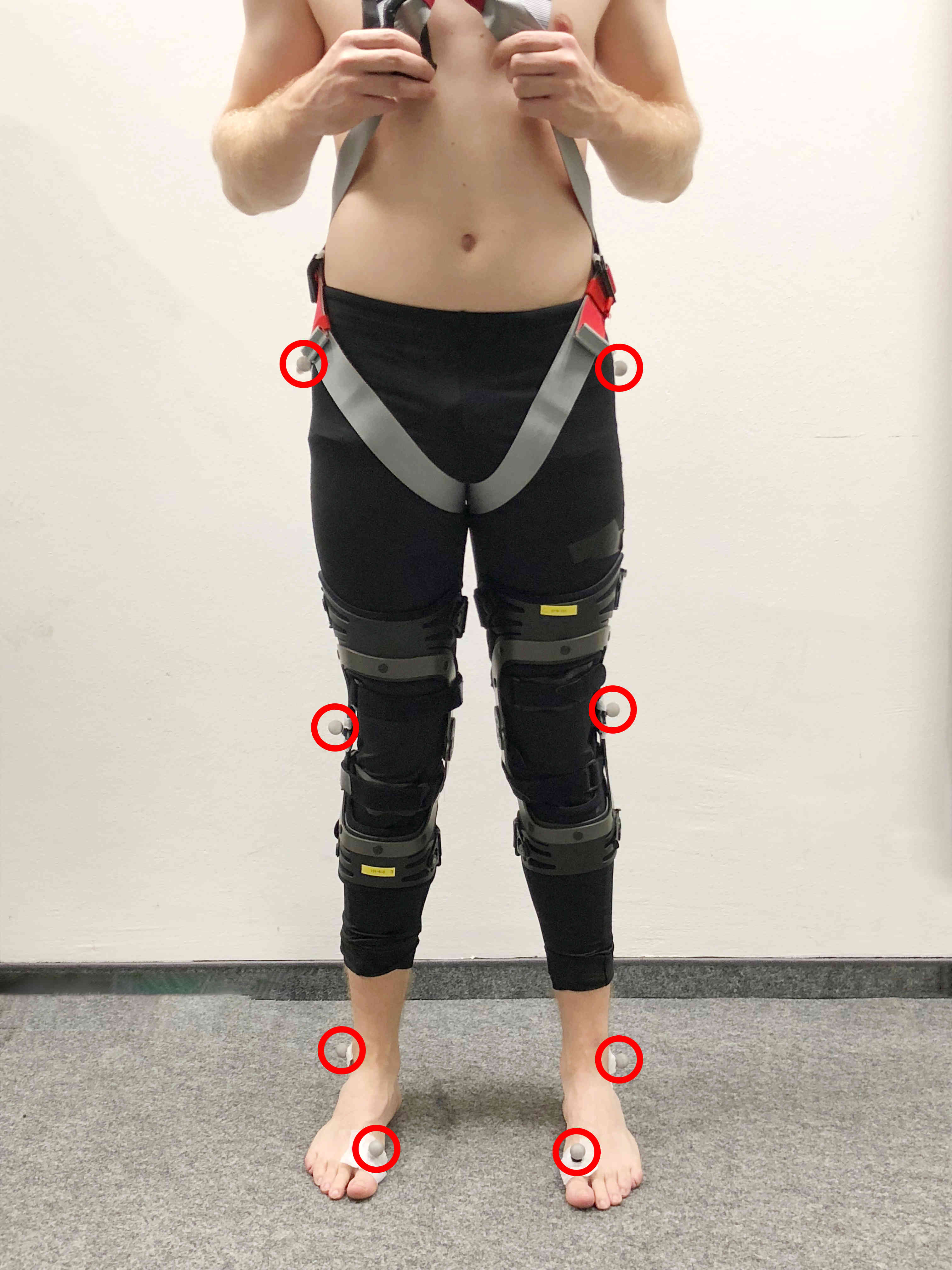}\hfill
    \includegraphics[width=0.49\columnwidth]{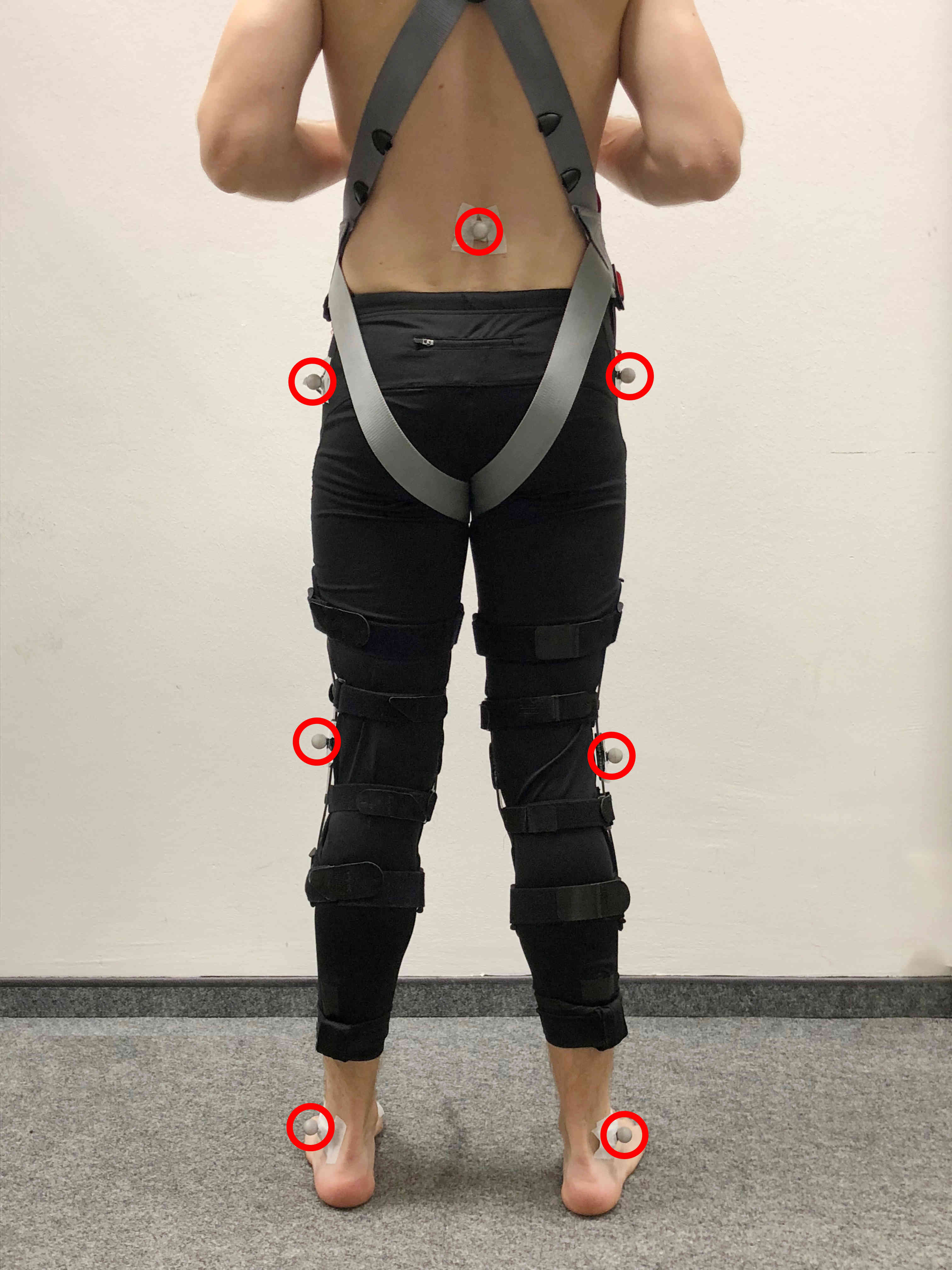}
    \caption{Placement of nine reflective markers (highlighted by red circles) on the subject for motion capturing.}
    \label{fig:mocap_setup}
\end{figure}

Vertical \acp{GRF} were individually recorded at $480$\,Hz for the left and right limb using a custom instrumented treadmill (type ADAL-WR, HEF Tecmachine, Andrezieux Boutheon, France). Based on integrated force-measuring plates (Kistler, Winterthur, Switzerland), one for each leg, the vertical \acp{GRF} (in $z$-direction) were acquired for the left and right leg separately.

A \ac{3D} motion capture system ($12$ high-speed infrared cameras, model Qqus, Qualisys, Sweden) recorded body segment kinematics from nine reflective markers at $500$\,Hz. \cref{fig:mocap_setup} shows the positioning of the reflective markers on a test subject for motion capturing. The nine markers were attached to the following body parts: toe (1st metatarsophalangeal joint), ankle (lateral malleolus), knee, hip (greater trochanter head), and lower back (sacrum). 

The mono-static radar systems measures the radial velocity of targets ($v_\text{rad}$), i.e., velocities along the propagation direction of the \ac{EM} wave. Since motion velocities along the $x$-direction are small and thus do not significantly contribute to the radial velocity component, we focus on motions in the sagittal plane, i.e., the $y$-$z$-plane, as illustrated in \cref{fig:exp_setup}. The radar is positioned at approximately knee height, such that for the knee's velocity we use $v_\text{rad} \approx v_{y,\text{moc}}$, where $v_{y,\text{moc}}$ is the velocity in $y$-direction measured by the motion capture system. The radial velocities of ankle and toe are approximated by $v_\text{rad} \approx v_{y,\text{moc}} \cos{\theta} + v_{z,\text{moc}} \sin{\theta} $, where $\theta = 18\degree$ for the presented setup. 

In order to simulate asymmetric gait, the right knee flexion was restricted by an adjustable orthosis (50K13 Knieorthese Genu Arexa, Otto Bock Healthcare GmbH, Duderstadt, Germany). Since wearing the orthosis changes the surface of the leg, and thus, its \ac{EM} reflection characteristics, the study participants wore a second, unconfined orthosis on the left leg to ensure comparability between the radar backscatterings from the left and the right leg. We note that the knee markers were attached to the pivot point of the orthoses' joint, since they covered the knee joints.

\subsection{Subject information and experimental protocol}
Nineteen individuals ($5$ females and $14$ males, aged $28.9 \pm 7.5$ years, height $176.5 \pm 10.2$\,cm, mass $72.3 \pm 11.5$\,kg) without gait impairment participated in the study. The study protocol was approved by the Technische Universit\"at Darmstadt ethics commission (EK 31/2018) and all volunteers provided written consent prior to participation. The experiments were conducted at the Locomotion Laboratory at Technische Universit\"at Darmstadt, Germany (no absorbers for \ac{EM} waves). The test subjects were asked to walk at the center of the treadmill for four consecutive minutes under five different conditions. For the first two minutes the treadmill speed was set to $0.7$\,m/s, and the speed switched automatically to $1.1$\,m/s for another two minutes. To account for speeding up and slowing down phases, the first and last $10$\,s of each two minute interval are neglected such that the measurement duration is $100$\,s for each walking speed. In the sequel, the two treadmill speeds $0.7$\,m/s and $1.1$\,m/s are referred to as \textit{slow} and \textit{fast} (speed), respectively. In a random order, the maximum knee flexion of the right knee was confined by an adjustable orthosis. Besides no confinement, which refers to normal gait, four different confinement angles, i.e., \ang{45}, \ang{30}, \ang{20}, \ang{10}, were investigated. Here, the angle describes the extent to which the knee could be bent, where \ang{0}~refers to a straight leg, e.g., while standing. These five experiments were performed twice in succession: once with the radar positioned in front of the test subject (five trials with randomized angle of deflection) and once behind the test subject (five trials with randomized angle of deflection). Between each experiment a rest period of approximately $5$\,min was required to adjust the orthosis and store the data. Due to synchronization errors between motion capture and \ac{GRF} data, there are only 18 samples available for normal walking and a knee angle confinement of \ang{45}.

\section{Methods}\label{sec:methods}
\begin{figure}
    \centering
    \includegraphics[clip,trim=2 0 0 0,width=0.85\columnwidth]{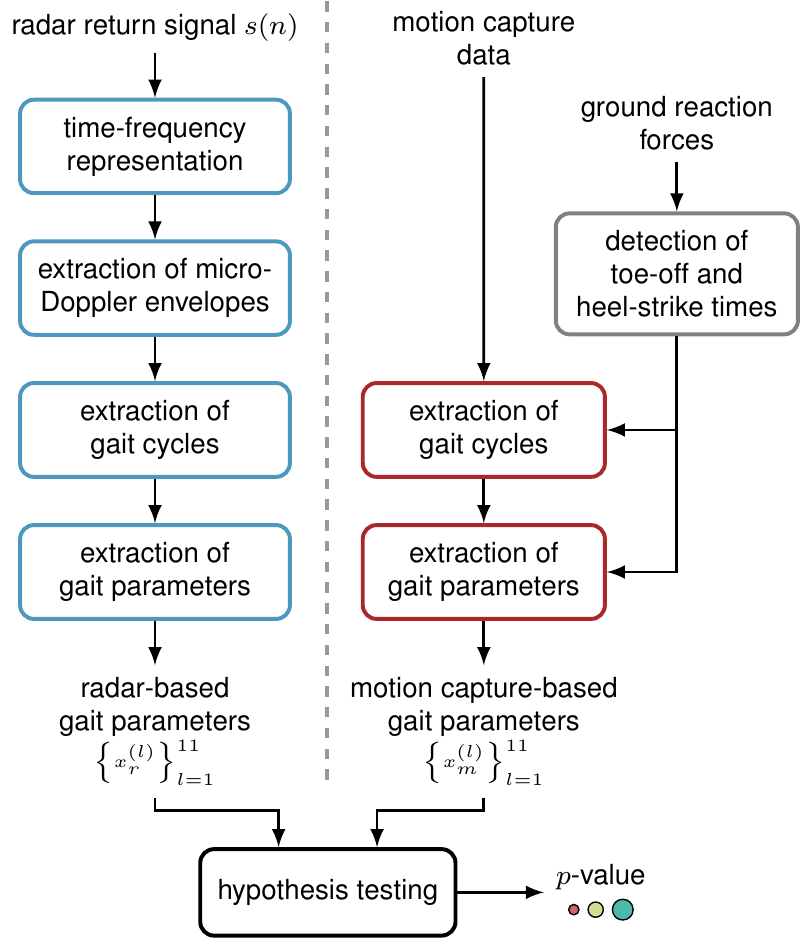}
    \caption{Flow diagram of processing steps for the extraction and comparison of eleven biomechanical parameters based on radar micro-Doppler signatures and motion capture data utilizing ground reaction forces.}
    \vspace{-0.5em}
    \label{fig:flowdiagram}
\end{figure}

\cref{fig:flowdiagram} gives an overview of the proposed framework for the extraction and comparison of gait parameters based on radar data as well as motion capture data and \acp{GRF}. In the following, the individual processing steps will be described in detail for both domains.

\subsection{Time-frequency representation of radar backscatterings}
\label{sec:radar_repr}

Since the radar backscatterings of gait motions are highly non-stationary, we analyze the data in the joint time-frequency domain \cite{Che11}. For human motion analysis, the spectrogram is typically utilized, which is obtained by the squared magnitude of the short-time Fourier transform, i.e., \cite{Opp99}
\begin{equation}\label{eq:spectrogram}
\mathrm{S}(n,k) = \left| \sum_{m=0}^{M-1} w(m) s(n+m) \exp{\left(-j 2 \pi \frac{m k}{M}\right)}\right|^2, 
\end{equation}
for $n = 0, \dots, N_r-1$, where $s(n)$ is the sampled zero-mean radar return of length $N_r$, $w(\cdot)$ is a smoothing window of length $M$, and $k$ is the discrete frequency index with ${k = 0, \dots, M-1}$. The radar data are recorded at a sampling frequency of $12.8$\,kHz and sub-sampled to $2.56$\,kHz. A Hamming window of length $M = 256$ is used with an overlap of $255$ samples, and zero-padding is applied to obtain $2048$ discrete frequency points \cite{Sei19}. Thus, we obtain a frequency resolution of $40$\,Hz (here equivalent to $0.25$\,m/s) and a time resolution of $0.1$\,s. For real time processing, the granularity along the time and frequency axis may be decreased to reduce computation time. Spectrograms of a person walking at $1.1$\,m/s on a treadmill are given in \cref{fig:specs_thr}, where in \cref{mDSig_front_env} the radar is positioned in front of, and in \cref{mDSig_back_env} behind the test subject. The ordinates of the spectrogram indicate both, the measured Doppler frequency $f_D$ and inferred radial velocity $v_\text{rad} $, which are related by
\begin{equation*}
    f_D \approx - f_t \frac{2 v_\text{rad} }{c}, \quad \text{for~} v \ll c,
\end{equation*}
where $f_t$ is the transmitting frequency of the radar and $c$ is the propagation speed of the \ac{EM} wave. The radial velocity $v_\text{rad}$ is defined in the direction of the propagation direction of the \ac{EM} wave and assumes positive values when the target is moving away from the radar. Hence, the observed Doppler shift is positive and negative for toward and away from radar motions, respectively (compare ordinates in \cref{mDSig_front_env,mDSig_back_env}). Here, we focus on those micro-Doppler components that arise from motions in walking direction, i.e., positive and negative Doppler shifts for toward and away from the radar, respectively. We note that the remaining micro-Doppler signatures are characteristic for treadmill walking and do not appear in this form in overground walking (see e.g.~\cite{Ami17, Sei17a}).

The background noise in the spectrogram is removed using an adaptive thresholding technique \cite{Kim09}. For each measurement, the noise level is estimated based on a frequency band without signal components. The amplitudes in the spectrogram are then lower bounded by this noise threshold. Based on the noise-reduced spectrogram, as shown in \cref{fig:specs_thr}, the micro-Doppler envelope signal is extracted. For each time instant, we detect the highest Doppler frequency, where significant signal energy is present. By use of morphological operations, as detailed in \cref{sec:mD_env}, it is ensured that relevant signal components are considered even if lower Doppler components have less energy at a particular time instant. Examples of envelope signals are shown in \cref{fig:specs_thr} by the black dashed lines.

To account for drifts in the sampling frequency of the systems, the motion capture and the radar data are synchronized prior to calculation of the spectrogram. For this, the micro-Doppler envelope is utilized. Analogously, a motion capture envelope is obtained by recording the maximal velocity among all recorded body parts for each time instant. Then, the maximum of the generalized cross-correlation function between the radar and motion capture envelope is used to synchronize the signals in time, such that their correlation is maximal. Next, minima in the radar and the motion capture envelope signal are detected and their pair-wise time difference is calculated. Given these time differences as a function of the measurement duration, a robust linear least squares fit is employed to determine the drift in sampling frequency of the radar system compared to the motion capture system. The average drift over all measurements is found as $0.01$\,s/sample, and used to re-sample each radar measurement prior to calculation of the spectrogram. The synchronization of the radar and motion capture data is also utilized to determine which micro-Doppler step signature corresponds to which leg. m

\subsection{Extraction of micro-Doppler envelope signals}
\label{sec:mD_env}

\begin{figure}[t!]
	\centering
	\subfloat[front view]{\includegraphics[clip, trim= 10 0 0 10,width=\columnwidth]{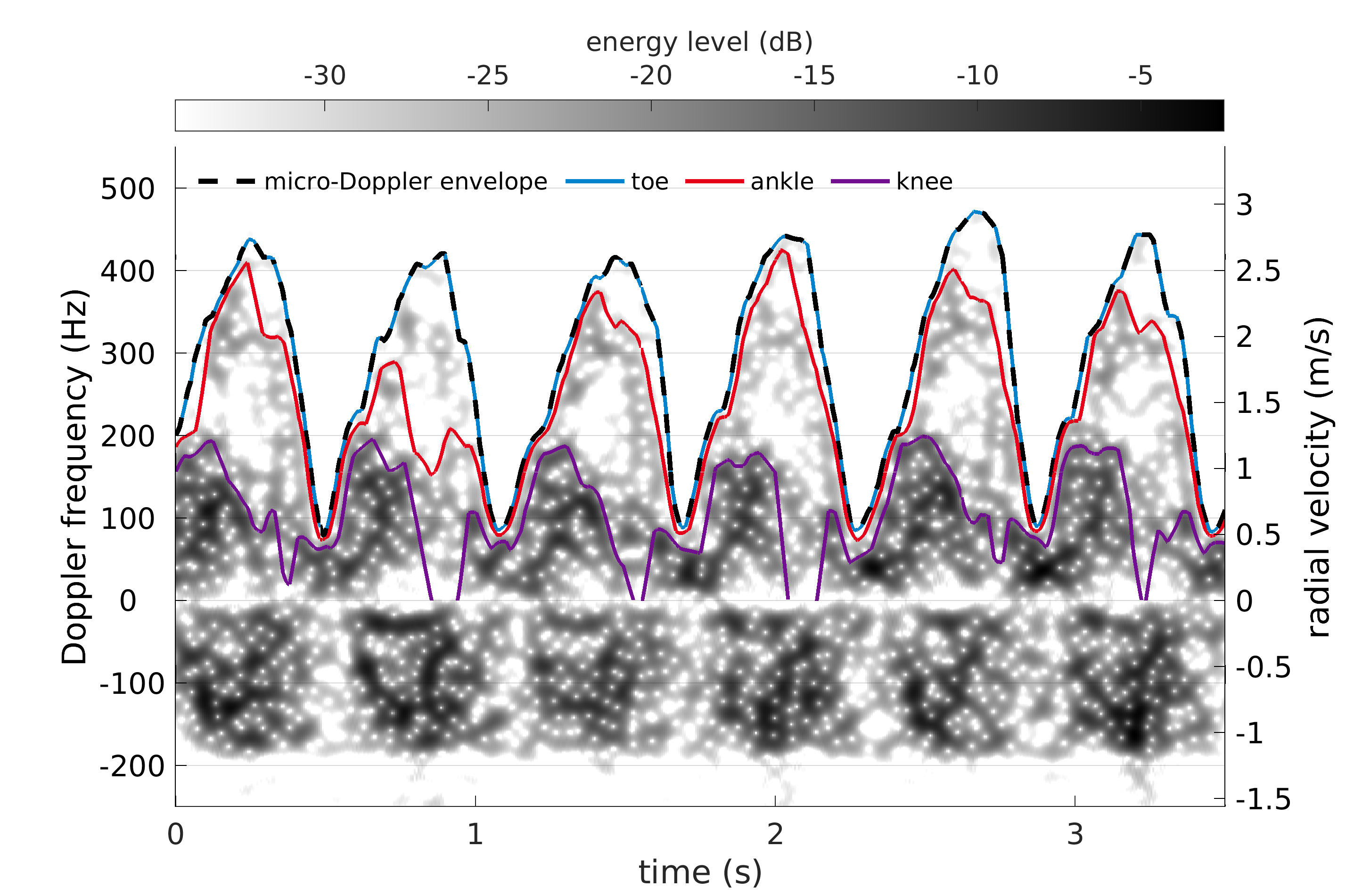}
		\label{mDSig_front_env}}\\ \vspace{-0.75em}
	\subfloat[back view]{\includegraphics[clip, trim= 10 0 0 10,width=\columnwidth]{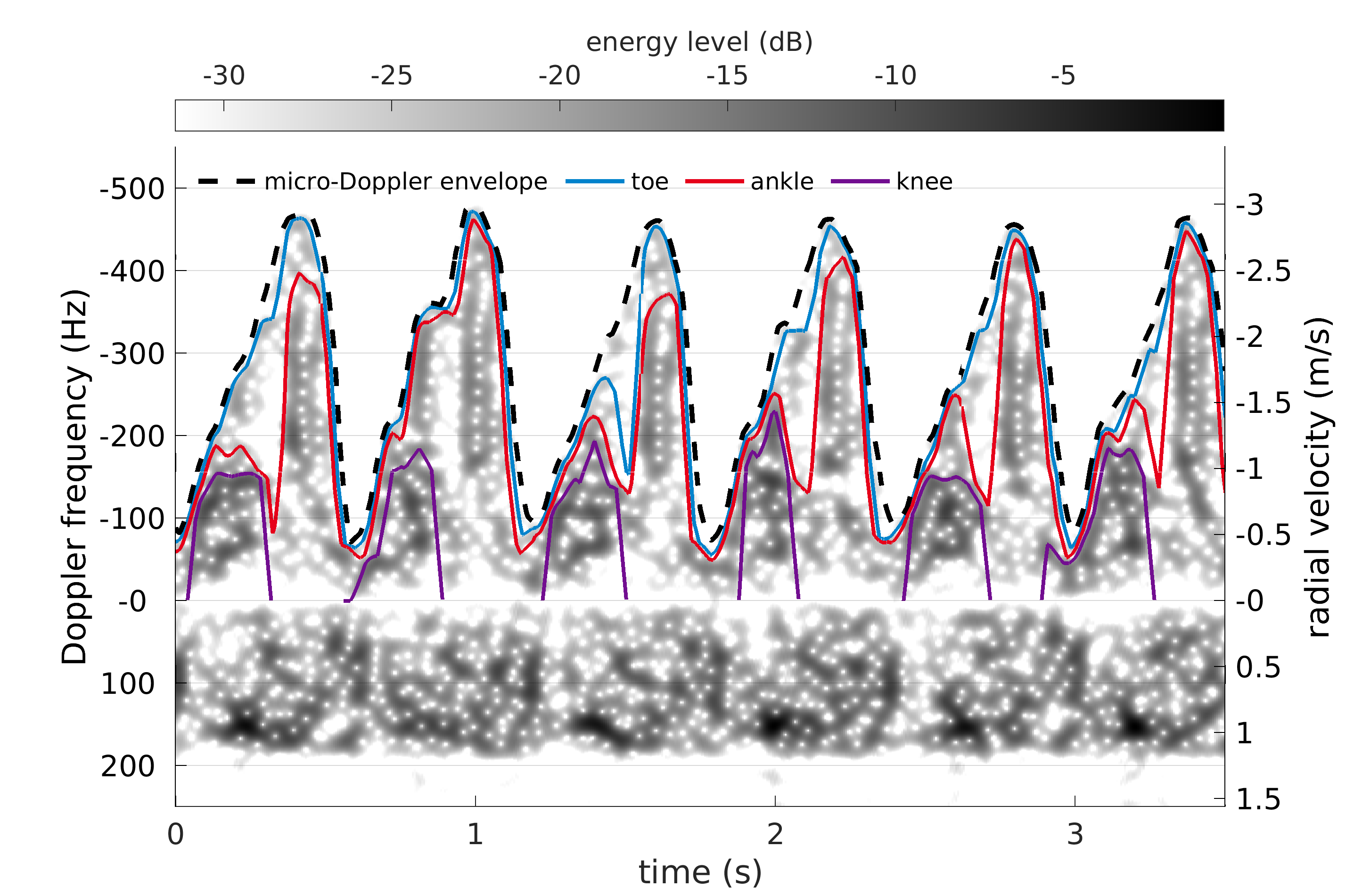}
		\label{mDSig_back_env}}
	\caption{Examples of micro-Doppler signatures. The black dashed lines are the standard micro-Doppler envelope signals, and the blue, red, and purple lines indicate the extracted envelope signals of toe, ankle and knee, respectively.}
	\label{fig:specs_thr}
\end{figure}

To obtain the standard micro-Doppler envelope signal, as shown in \cref{fig:specs_thr} by the dashed black lines, a series of morphological operations are applied to the spectrogram. First, the spectrogram as given by \cref{eq:spectrogram} is converted to a binary image, where values below the estimated noise threshold are set to zero. Next, the binary image is dilated using a rectangular structuring element of size $16 \times 32$, where $16$ pixels correspond to approximately $20$\,Hz along the Doppler frequency axis and $32$ pixels represent $12.5$\,ms along the time axis. After filling holes in the resulting image, unconnected components with less than \num{40000} pixels are removed. Finally, the image is eroded twice using a diamond-shaped structuring element with an extension of $8$ pixels in each dimension and afterwards unconnected components with less than $500$ pixels are removed. Using this image, the first index where signal energy occurs is stored for each time instant, which results in the micro-Doppler envelope signal.

The standard micro-Doppler envelope signal is frequently used to determine the maximal velocity of the swinging feet \cite{Sei19}. Here, we utilize the above described procedure to also extract the maximal velocity of other parts of the lower limbs, i.e., ankle and knee. To do so, we adjust the noise threshold such that low energy components are neglected. The intuition is that the radar backscatterings from the ankle and the knee are expected to be stronger compared to those of the toe. Thresholding the micro-Doppler signatures using a proportionally higher noise threshold results in micro-Doppler envelope signals that capture the ankle and knee velocities throughout a gait cycle. Examples of the extracted toe, ankle and knee envelope signals are shown in \cref{fig:specs_thr}. The toe envelope signal is obtained using the adaptive noise threshold and $85$\% of it for the front and back view, respectively.
Using $85\%$ and $75\%$ of the adaptive noise threshold for the front and back radar, respectively, the ankle envelope signal is extracted. To extract the knee envelope signal, the spectrogram is thresholded at $55\%$ of the adaptive noise threshold for both radar perspectives. For further processing, the envelope signals are averaged over all gait cycles in a measurement. Examples of averaged micro-Doppler envelope signals for different lower limb joint are shown in \cref{rad_env_normal_fast_front,rad_env_abnormal_fast_front,rad_env_normal_slow_front,rad_env_normal_fast_back}.

\subsection{Extraction of gait cycles}
\label{sec:grfmocap}

\begin{figure}[t!]
    \centering
    \includegraphics[clip,trim=0 0 10 10,width=0.8\columnwidth]{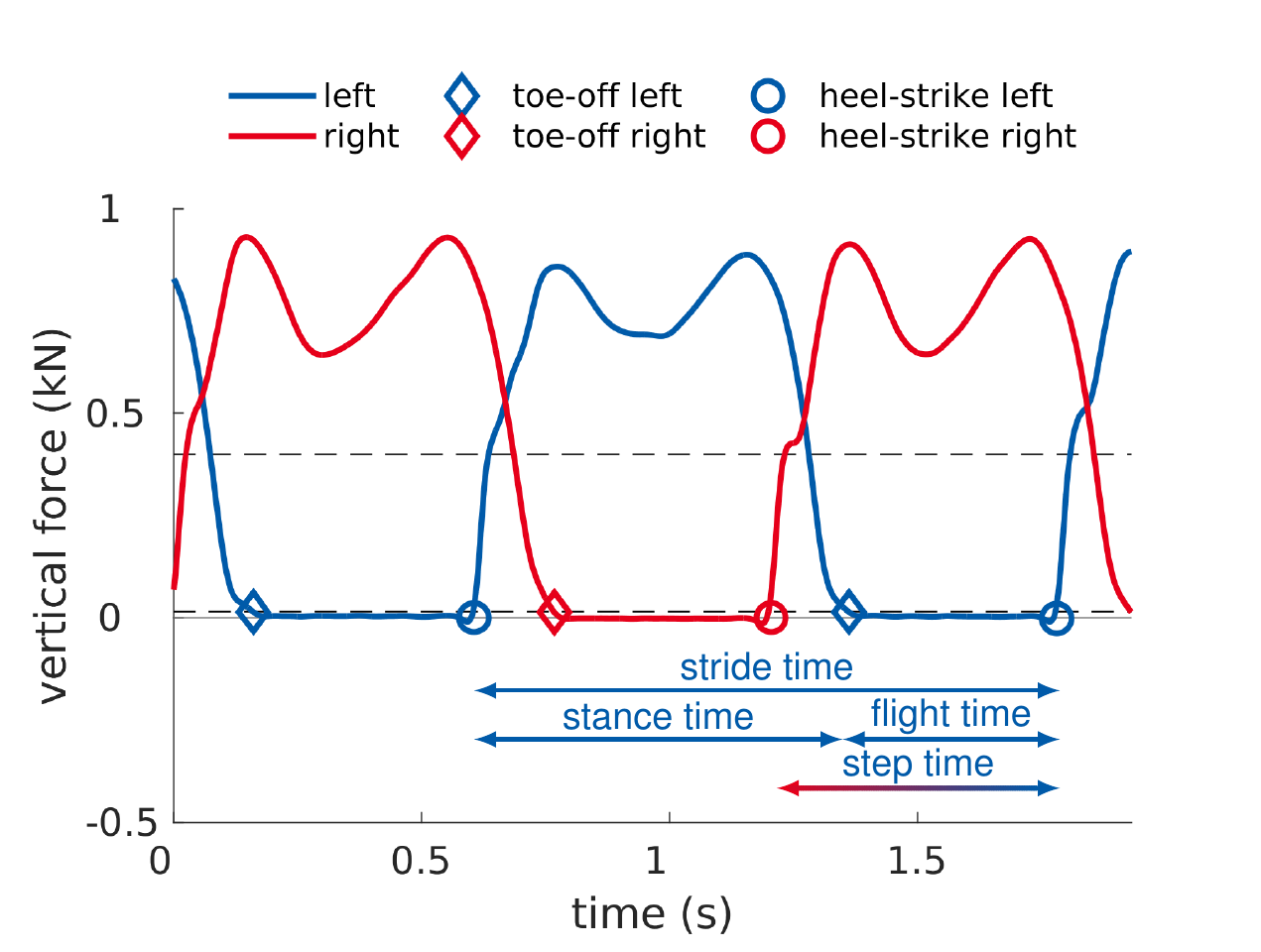}
    \caption{Example of vertical \acp{GRF} at $1.1$\,m/s, where diamonds and circles mark the detected toe-off and heel-strike events, respectively. The dashed lines mark the thresholds used for toe-off and heel-strike detection. For the left foot, the stride time, stance time, flight time and step time are indicated.}
    \label{fig:grf}
    \vspace{-0.5em}
\end{figure}

Vertical \acp{GRF} are used to determine the toe-off and heel-strike times of each stride. First, the vertical \ac{GRF} data are up-sampled to $1$\,kHz. In order to determine heel-strike events, a threshold of $400$\,N is employed \cite{Gri19}. Once the \ac{GRF} exceeds this threshold, the heel-strike time is found by the time instant of the last preceding force value that falls below a threshold of $15$\,N, and is either greater than its subsequent force value or smaller than $1$\,N. To detect toe-off times, the \acp{GRF} are filtered with a zero-lag, second-order, low-pass Butterworth filter (cutoff frequency $20$\,Hz). Then, starting $100$\,ms after each heel-strike time, the first \ac{GRF} value that falls below a threshold of $15$\,N is used to determine the toe-off time. \cref{fig:grf} shows an example of vertical \ac{GRF} for both legs at a treadmill speed of $1.1$\,m/s, where diamonds and circles indicate the toe-off and heel-strike times, respectively. A full gait cycle is defined from one heel-strike of one leg to the next one of the same leg. The duration of a gait cycle constitutes the stride time. The stance time for one leg is given by the duration from heel-strike to toe-off time, where the flight time is determined by the time from toe-off to the next heel-strike. The step time is defined as the time between the heel-strike of one leg until the next heel-strike of the other leg. As indicated in \cref{fig:grf}, for the left leg, we define the step time as the duration from the heel-strike of the right leg to the heel-strike of the left leg.

\begin{figure*}[t!]
  	\centering{
		\subfloat[mocap: normal, $1.1$\,m/s, front]{\includegraphics[clip, trim= 0 22 8 0, height=4.2cm]{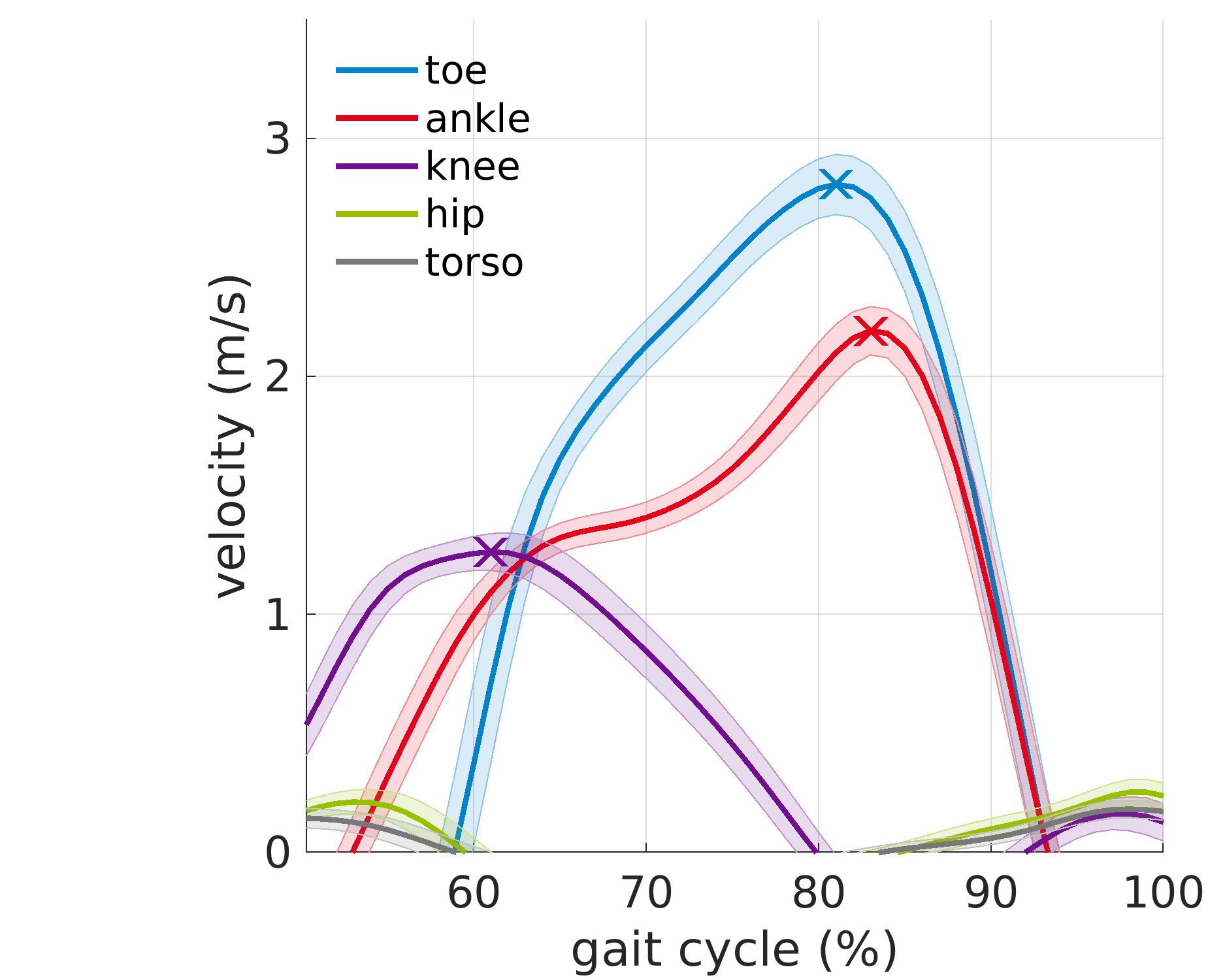}%
			\label{moc_normal_fast_front}}\hfill
		\subfloat[mocap: $10$\textdegree, $1.1$\,m/s, front]{\includegraphics[clip, trim= 110 22 8 0, height=4.2cm]{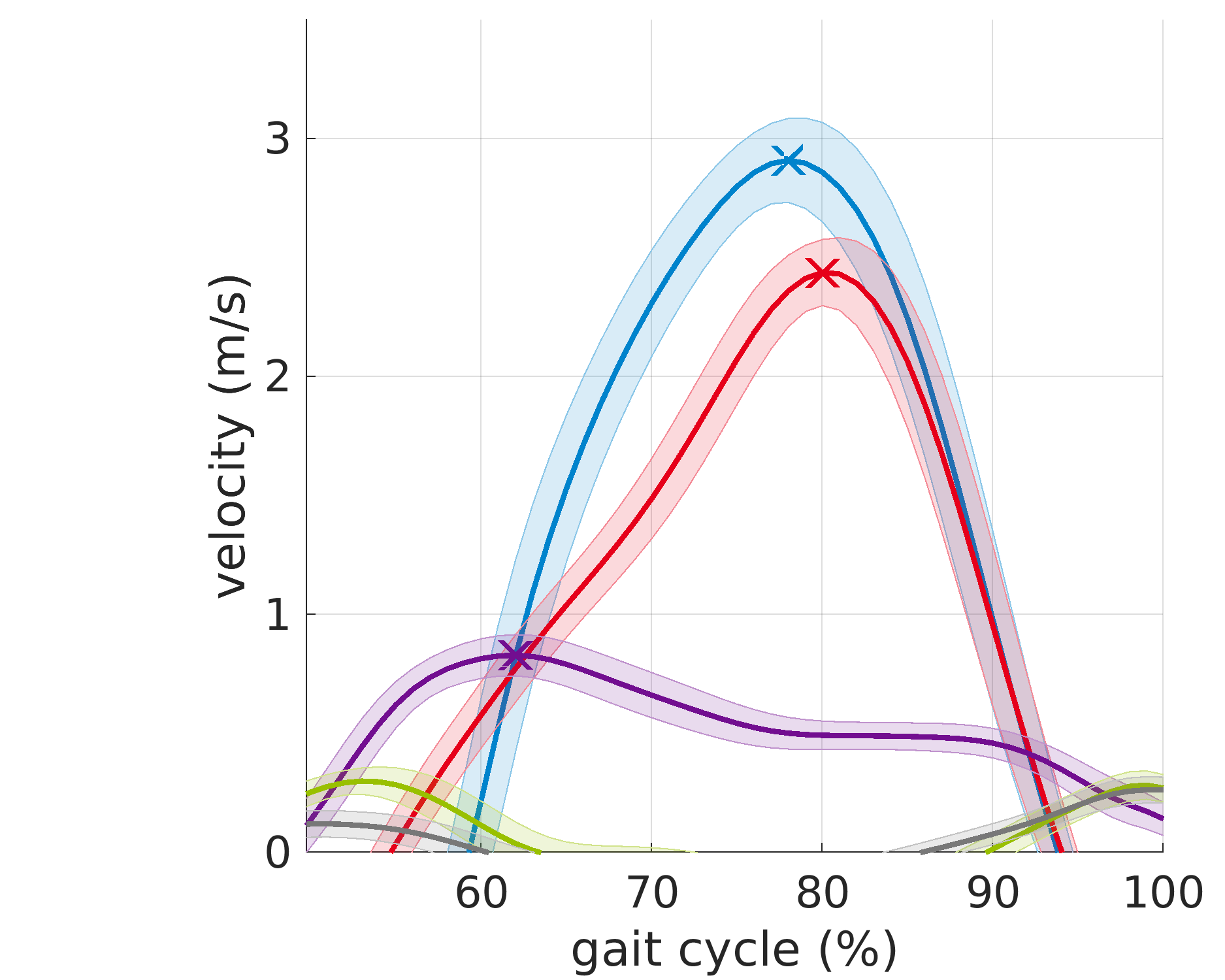}%
			\label{moc_abnormal_fast_front}}\hfill
		\subfloat[mocap: normal, $0.7$\,m/s, front]{\includegraphics[clip, trim= 110 22 8 0, height=4.2cm]{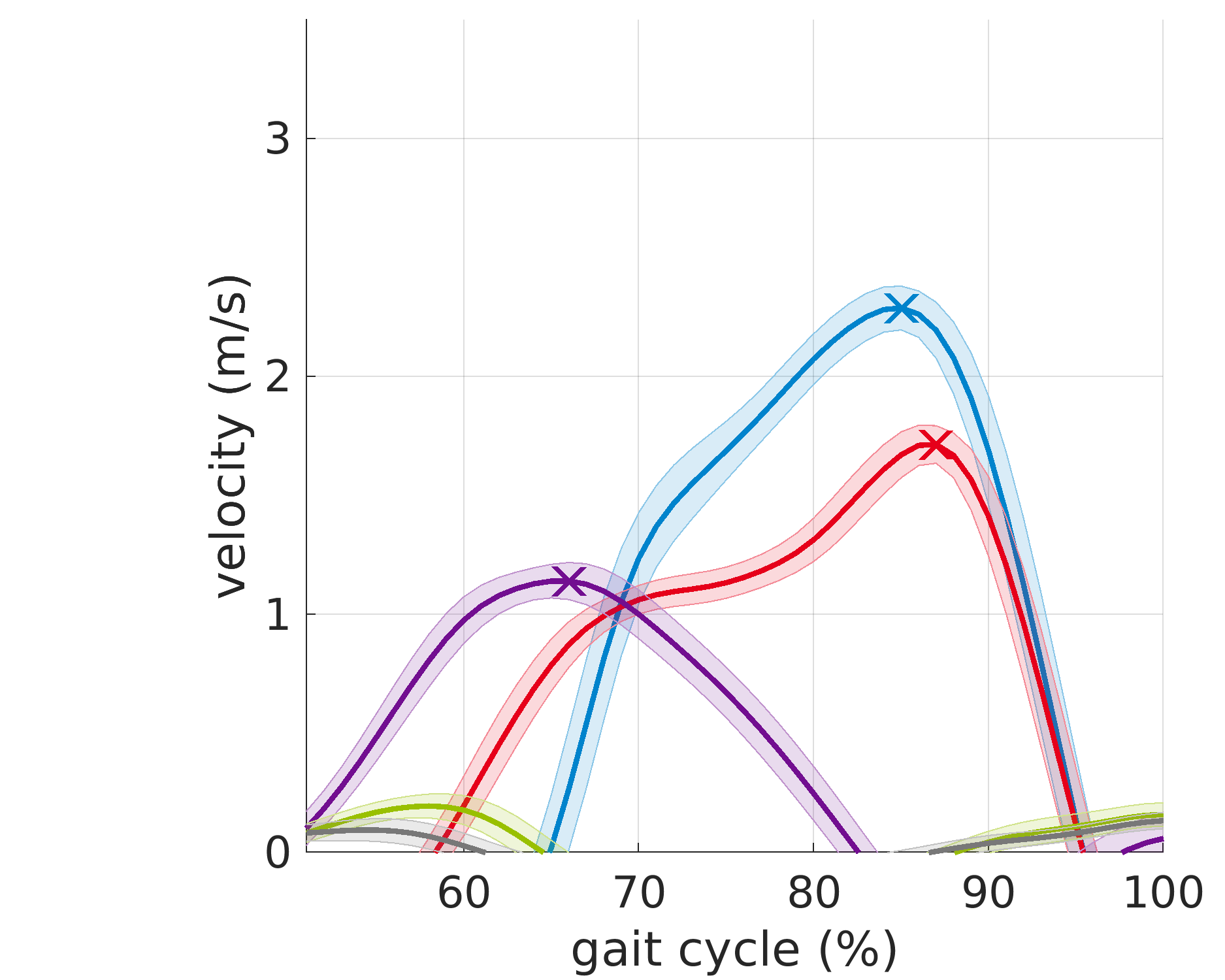}%
			\label{moc_normal_slow_front}}\hfill
		\subfloat[mocap: normal, $1.1$\,m/s, back]{\includegraphics[clip, trim= 110 22 8 0, height=4.2cm]{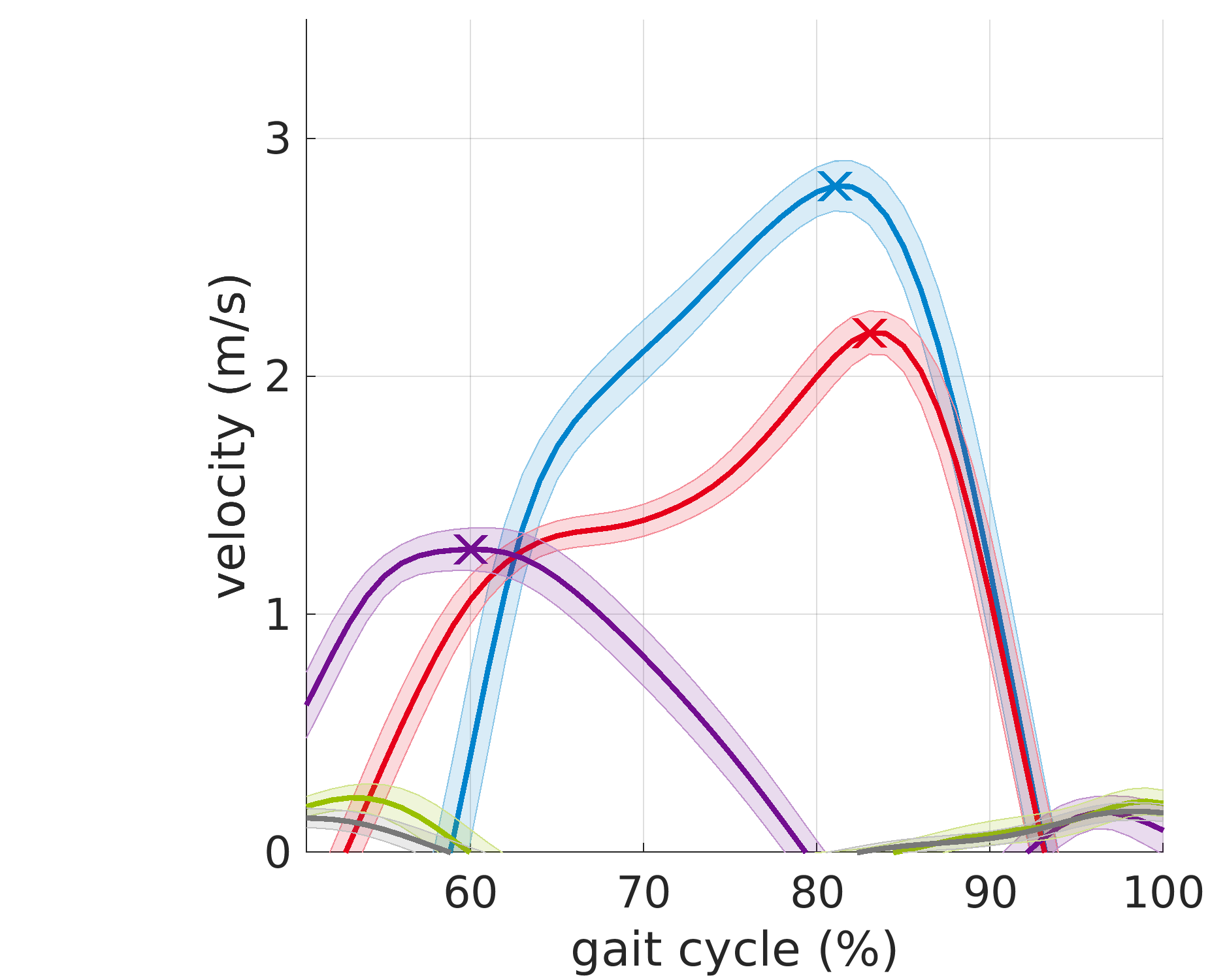}%
			\label{moc_normal_fast_back}}}\vspace{-0.8em}
  	\centering{
  		\subfloat[radar: normal, $1.1$\,m/s, front]{\includegraphics[clip,trim= 0 27 8 0, height=4.23cm]{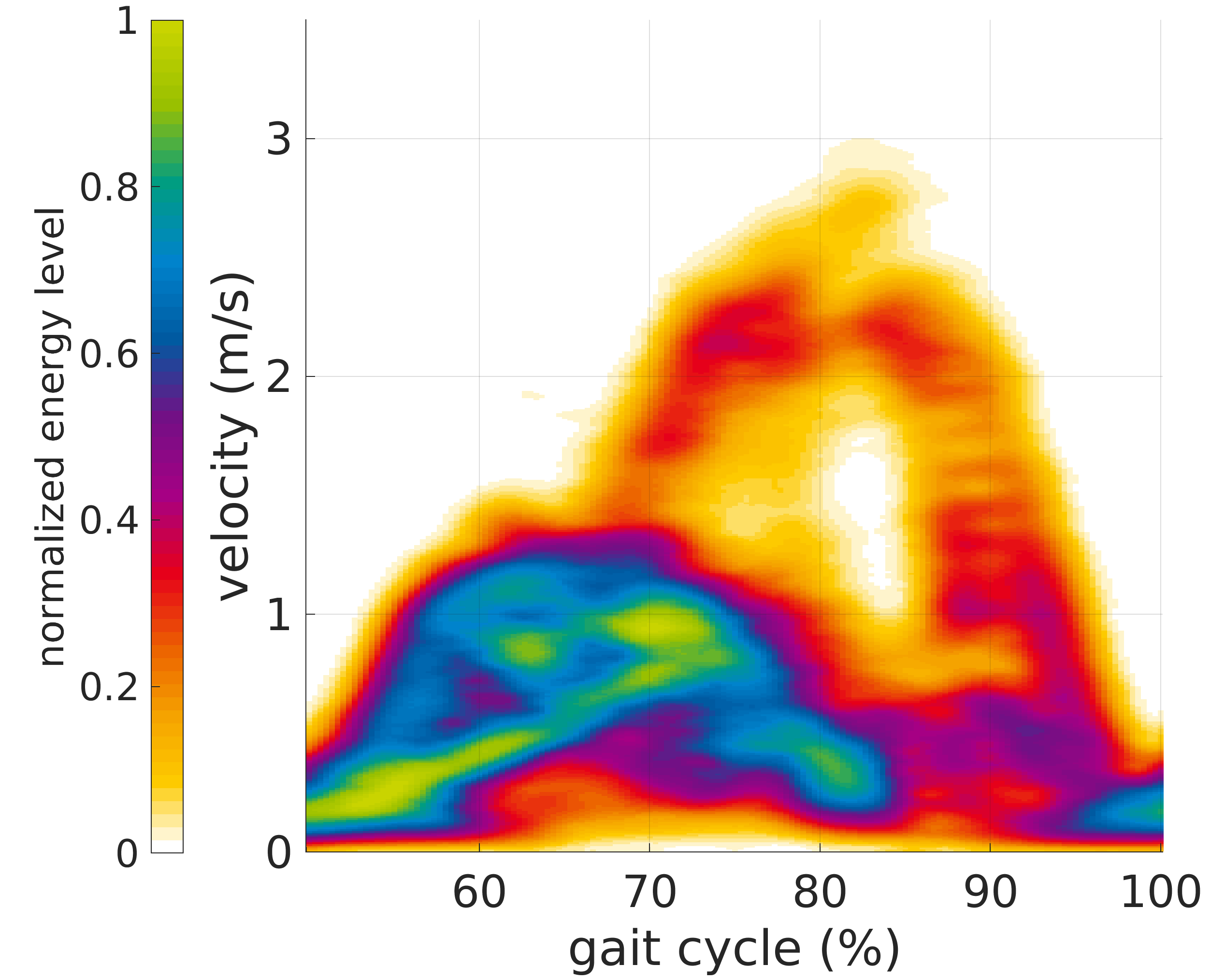}%
			\label{rad_normal_fast_front}}\hfill
		\subfloat[radar: $10$\textdegree, $1.1$\,m/s, front]{\includegraphics[clip,trim= 146 27 8 0, height=4.23cm]{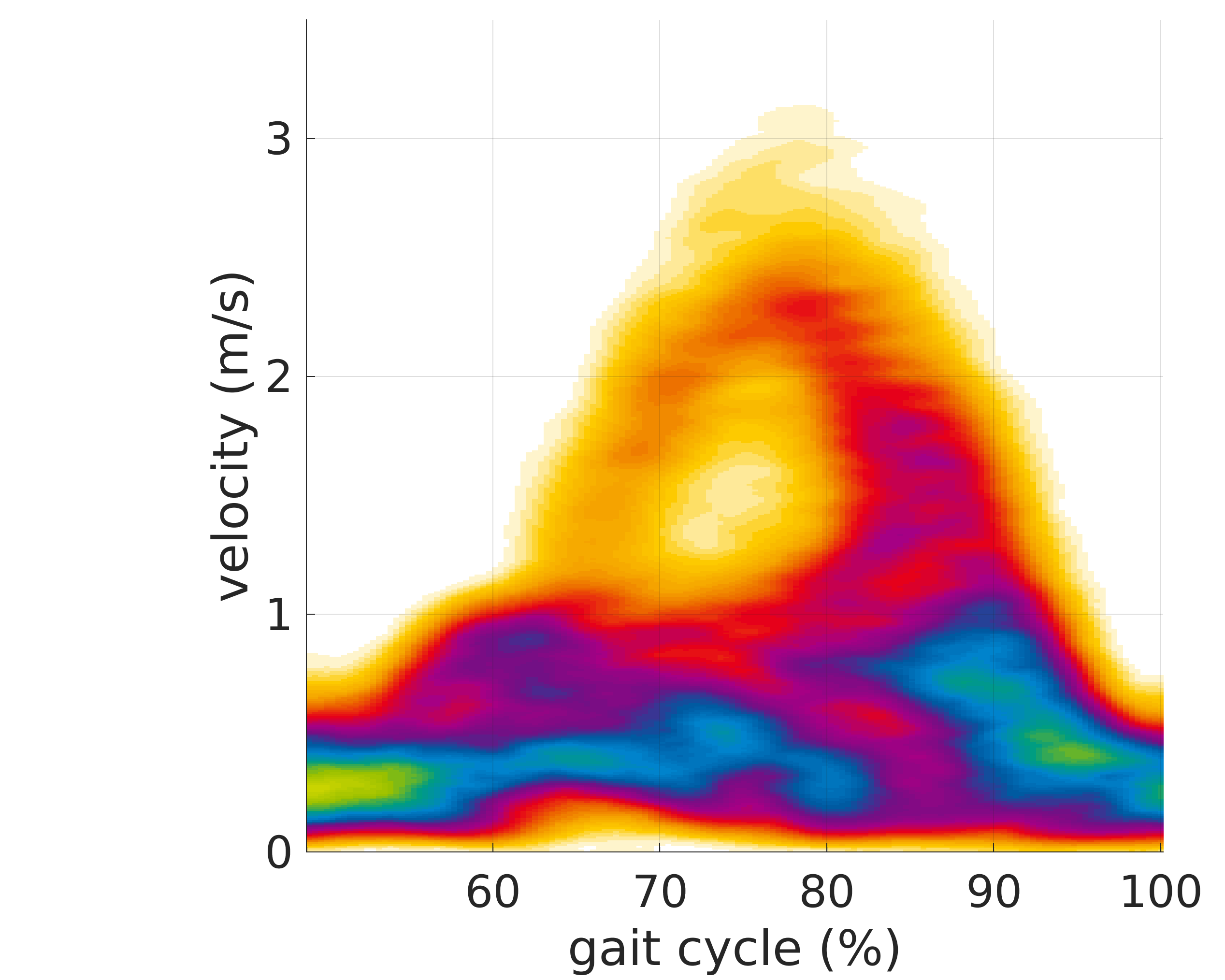}%
			\label{rad_abnormal_fast_front}}\hfill
		\subfloat[radar: normal, $0.7$\,m/s, front]{\includegraphics[clip,trim= 146 27 8 0, height=4.23cm]{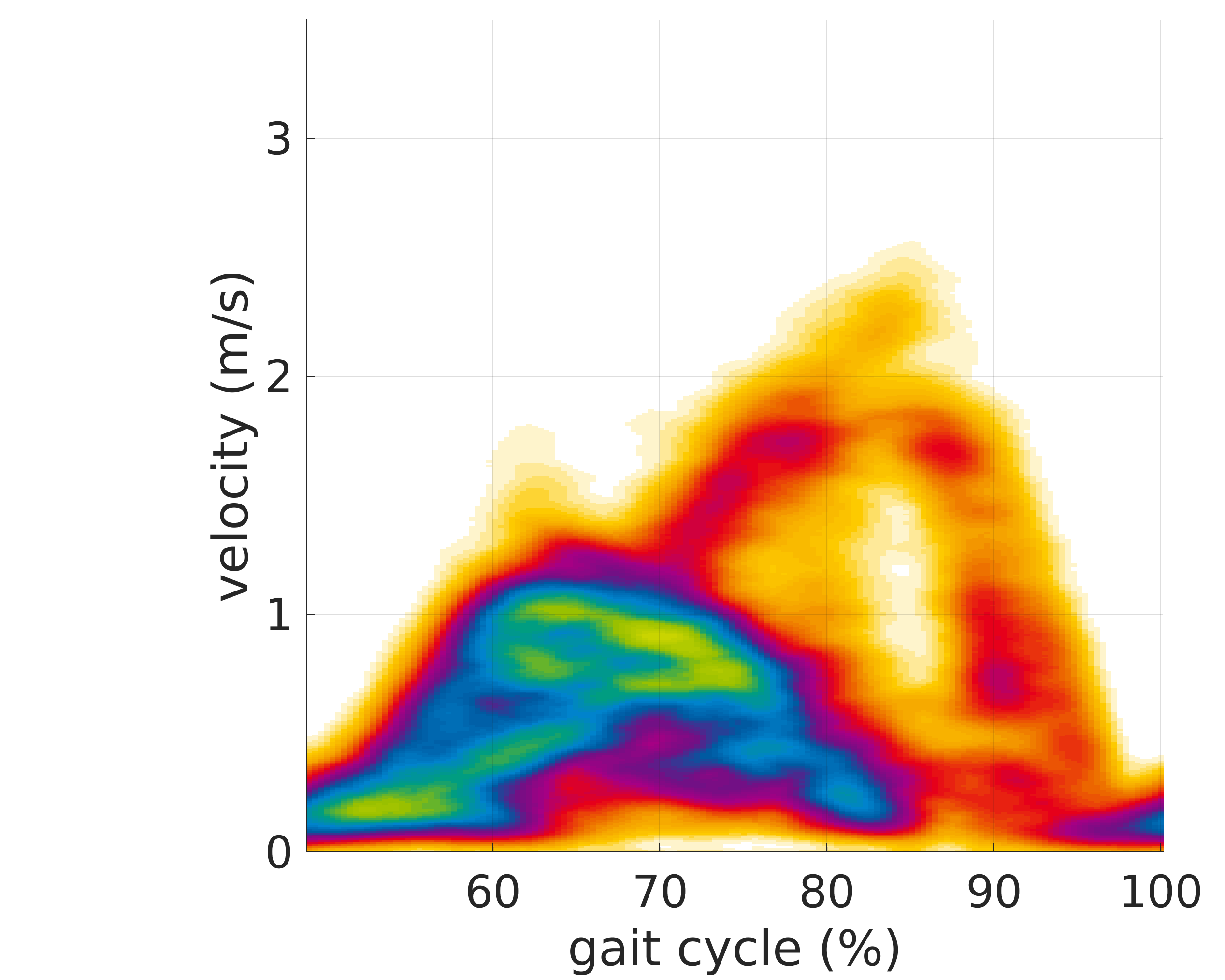}%
			\label{rad_normal_slow_front}}\hfill
		\subfloat[radar: normal, $1.1$\,m/s, back]{\includegraphics[clip,trim= 146 27 8 0, height=4.23cm]{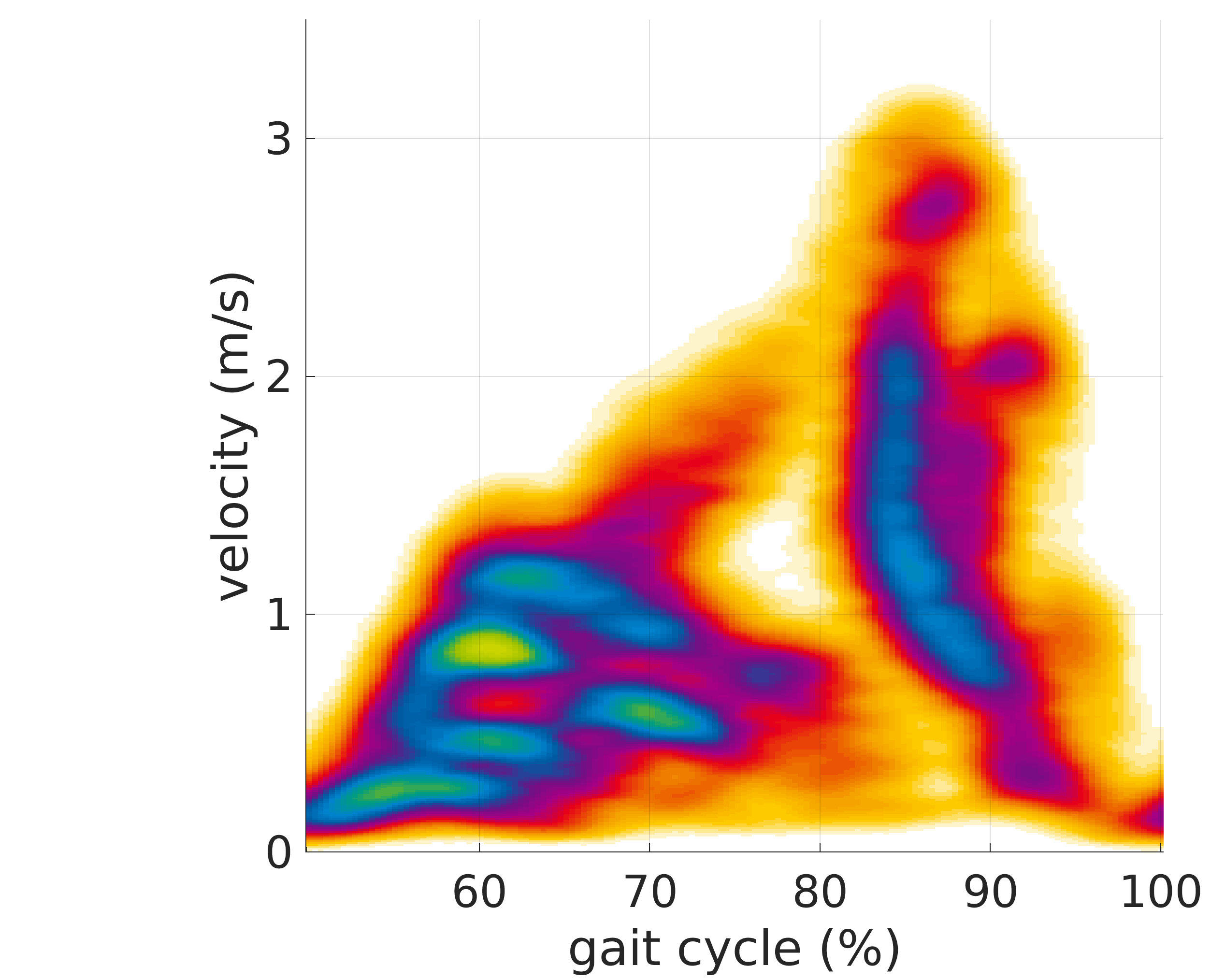}%
			\label{rad_normal_fast_back}}}\vspace{-0.8em}
  	\centering{
  		\subfloat[radar: normal, $1.1$\,m/s, front]{\includegraphics[clip,trim= 0 0 8 0, height=4.48cm]{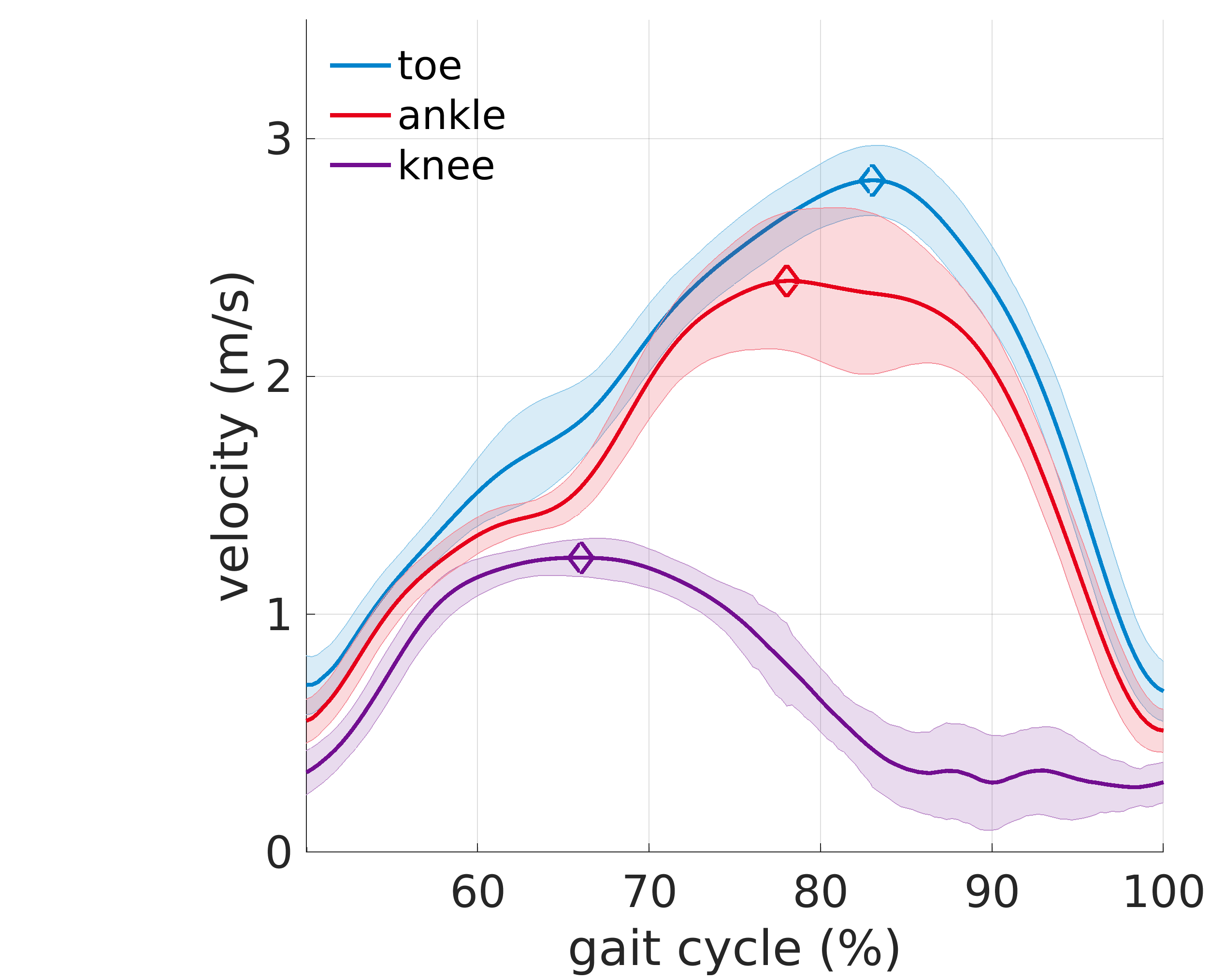}%
			\label{rad_env_normal_fast_front}}\hfill
		\subfloat[radar: $10$\textdegree, $1.1$\,m/s, front]{\includegraphics[clip,trim= 146 0 8 0, height=4.48cm]{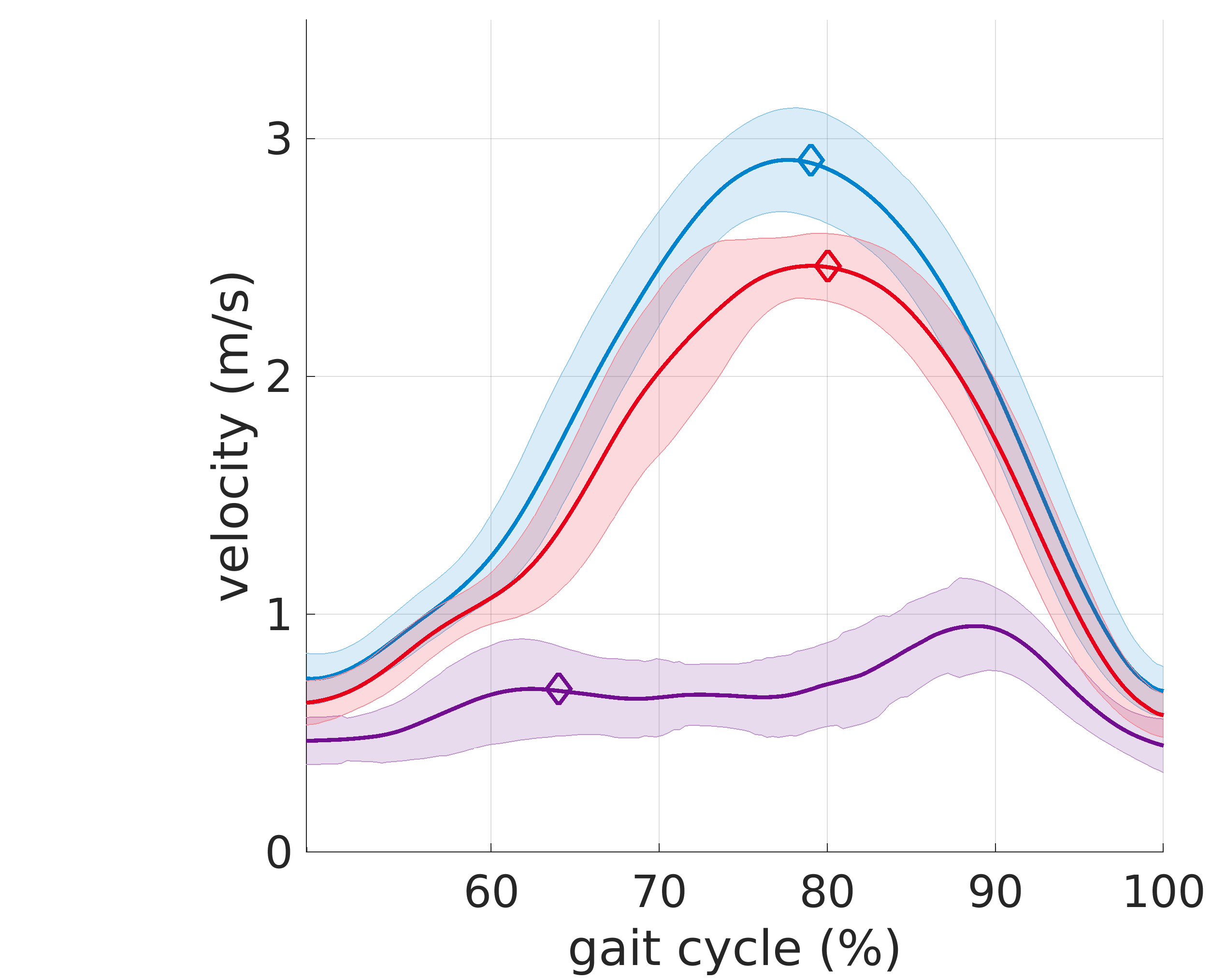}%
			\label{rad_env_abnormal_fast_front}}\hfill
		\subfloat[radar: normal, $0.7$\,m/s, front]{\includegraphics[clip,trim= 146 0 8 0, height=4.48cm]{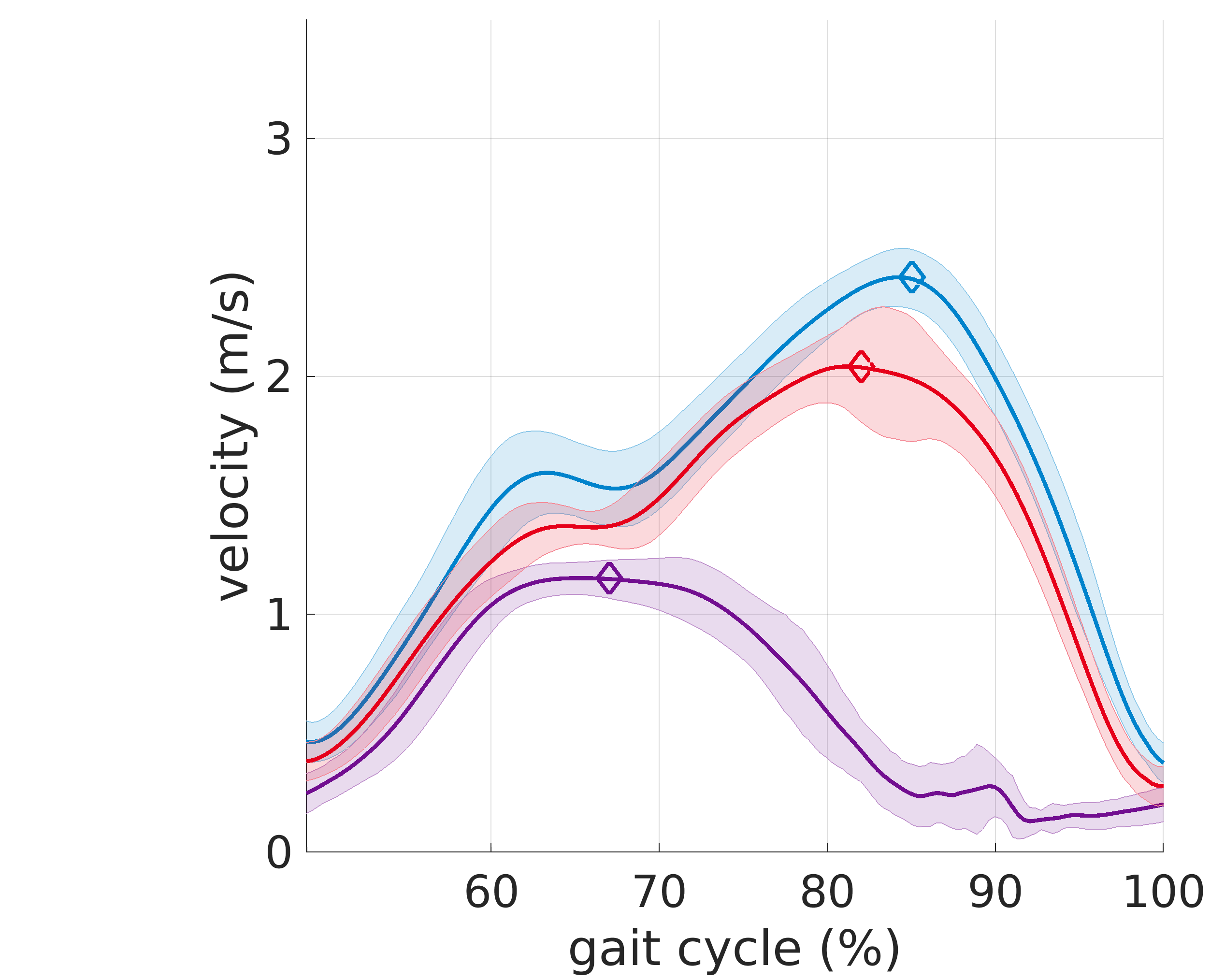}%
			\label{rad_env_normal_slow_front}}\hfill
		\subfloat[radar: normal, $1.1$\,m/s, back]{\includegraphics[clip,trim= 146 0 8 0, height=4.48cm]{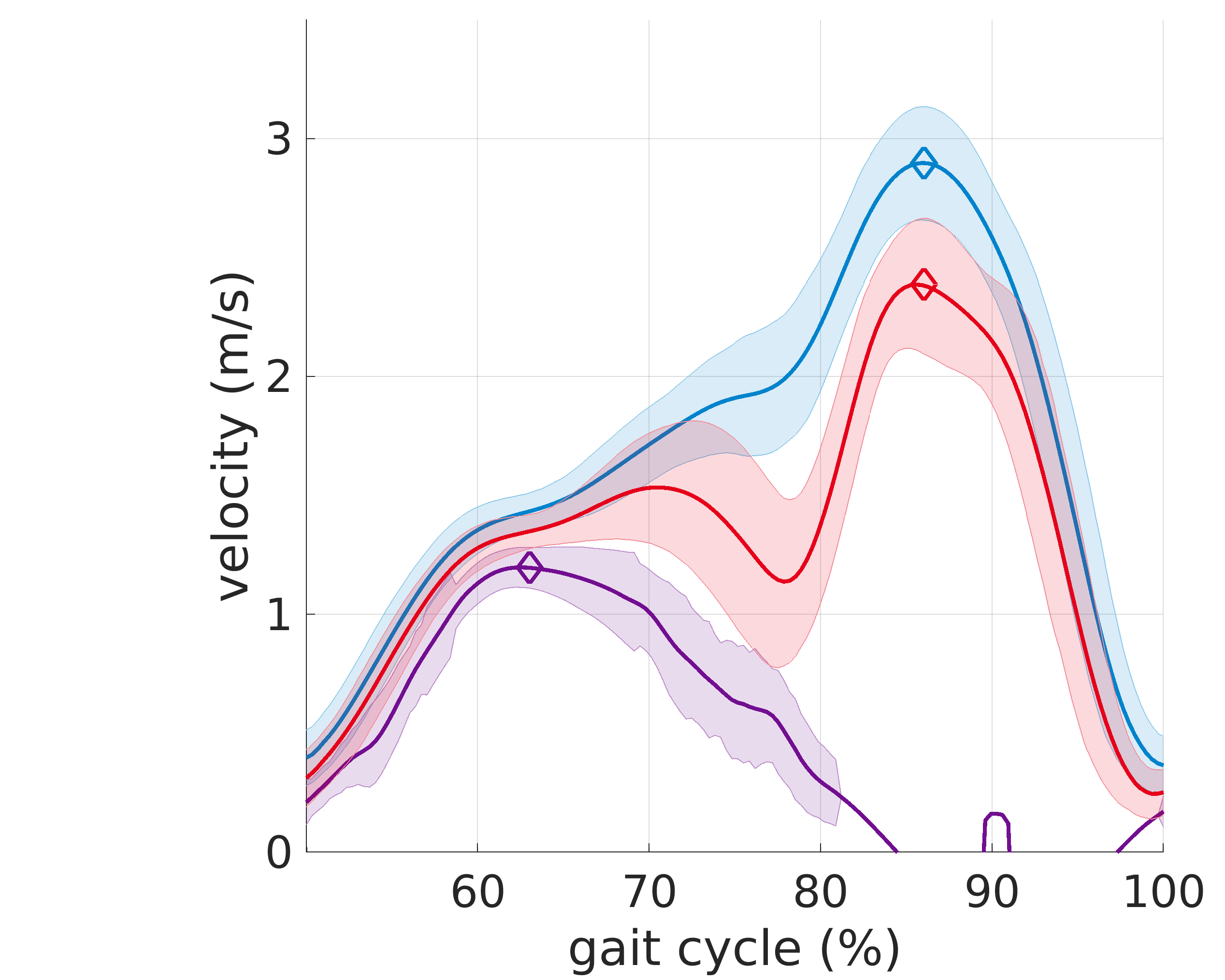}%
			\label{rad_env_normal_fast_back}}}
    \caption{Motion capture (first row) and radar (second and third row) data of the right leg for one individual. The four columns represent the analyzed conditions: degree of knee angle confinement (no restriction vs.~\ang{10}), walking speed ($1.1$\,m/s vs.~$0.7$\,m/s), and radar positioning (front vs.~back view). The time frame shown is the step time. In the first and third row, shaded areas indicate one standard deviation from the mean, and crosses and diamonds mark the maximal velocities of the lower limb joints for motion capture and radar data, respectively. The second row shows the averaged radar micro-Doppler signatures.}
    \label{fig:conditions_moc_rad}
    \vspace{-0.5em}
\end{figure*}

In order to match the frequency of the \ac{GRF} data, the motion capture data are up-sampled to $1$\,kHz. The raw marker data are filtered using a zero lag, fourth-order, low-pass Butterworth filter (cutoff frequency $10$\,Hz). Then, \acp{GRF} of left and right leg are utilized to cut the motion capture data into individual gait cycles. The velocity information of each marker is averaged over all gait cycles in a measurement. The portions of the gait cycle that represent the step time of the right leg are shown in \cref{moc_normal_fast_front,moc_abnormal_fast_front,moc_normal_slow_front,moc_normal_fast_back}. Hip and torso velocities are small since the test subjects were walking on a treadmill. The knee velocity generally reaches its maximum earlier in the gait cycle than the toe and ankle velocities. The latter exhibit higher maximal velocities compared to the knee. 

To extract portions of the spectrogram that correspond to one gait cycle, the standard micro-Doppler envelop is utilized (see \cref{fig:specs_thr}). Here, every second minimum in the micro-Doppler envelope signal defines the onset of a gait cycle for a particular leg. The so obtained micro-Doppler gait cycle signatures are averaged for each measurement. The resulting micro-Doppler signatures are less distorted and easier to interpret (visually). Examples of averaged micro-Doppler signatures for the right leg are shown in \cref{rad_normal_fast_front,rad_abnormal_fast_front,rad_normal_slow_front,rad_normal_fast_back}.

\begin{figure}[t!]
	\centering
	\includegraphics[clip,trim=0 0 10 10, width=0.75\columnwidth]{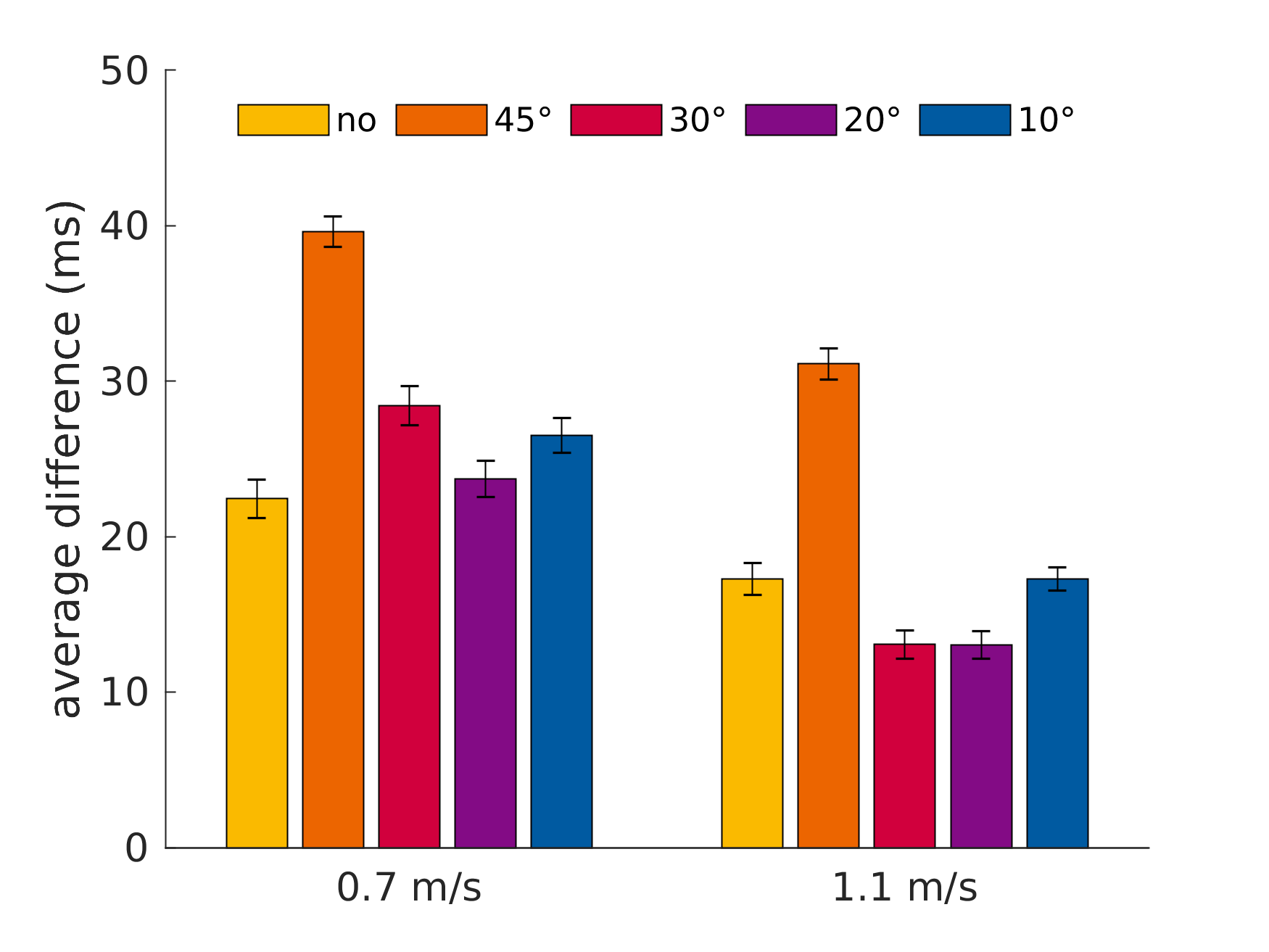}
	\caption{Average differences between the time instant of maximal knee velocity from the motion capture data and the toe-off time from the \acp{GRF} for five different knee angle confinements. The error bars indicate the $95$\% confidence interval for the average differences, where the Gaussian assumption on the distribution of the average difference was verified empirically.}
	\label{fig:flighttime}
	\vspace{-0.6em}
\end{figure}

\subsection{Definition of flight time}
\label{sec:flight_time_definition}

The flight time of a foot is defined by the time duration between the toe-off time and the subsequent heel-strike of the same foot \cite{Gou17}. Since the radar system records the radial velocities of all body parts simultaneously, the flight time of a foot is generally not directly accessible from the radar micro-Doppler signatures, i.e., in the joint time-frequency domain, the time-varying Doppler shifts of a foot are obscured by those of other body parts. In general, it is a non-trivial task to separate these components \cite{Abd17}. Here, we propose to use the time instant at which the knee reaches its maximal velocity to indicate the beginning of the flight time, i.e., the toe-off time. 

In order to justify this approach we show that the time instant of maximal knee velocity $t^{\text{moc}}_\text{knee-max-v}$ serves as a good approximation for the toe-off time $t^{\text{grf}}_\text{toe-off}$, where the former is obtained from motion capture data and the latter is determined based on vertical \acp{GRF}.
For this, we analyze the time difference $\Delta t = t^{\text{frc}}_\text{toe-off} - t^{\text{moc}}_\text{knee-max-v}$, which is extracted for each gait cycle. \cref{fig:flighttime} shows the average differences for the right leg, where the error bars indicate the $95$\% confidence interval for the mean. For both, the slow and the fast walking speed, and for all conditions the average difference between the time instant of the maximal knee velocity and the toe-off time is small. The average differences range between $13$ and $40$\,ms. Given the average flight time assumes $448$\,ms and $402$\,ms for $0.7$\,m/s and $1.1$\,m/s, respectively, the maximally expected error in flight time is approximately $9$\% ($40$\,m/s) and $8$\% ($31$\,m/s).

\subsection{Extraction of gait parameters}
\label{sec:gait_param_extr}

In total, eleven gait parameters are considered. Where applicable, the kinematic meaning of each of these parameters is given first \cite{Gou17}. Then the extraction methods for \ac{GRF} or motion capture and radar data are explained.

\subsubsection{Stride time}
A full gait cycle is defined from one heel-strike to the next. The duration of a gait cycle is given by the stride time. Utilizing \acp{GRF}, the average time-span between two consecutive heel-strikes of the same leg yields the stride time. In case of radar data, every second minimum in the micro-Doppler envelope signal indicates the beginning of a new gait cycle. Thus, the average time between every second minimum serves as an estimate for the stride time.

\subsubsection{Stance time}
The stance time refers to the time duration that a foot is on the ground during a gait cycle. From \ac{GRF}, the stance time can be obtained as the average time between a heel-strike and subsequent toe-off event. Given radar data, the stance time is calculated by subtracting the flight time from the stride time.

\subsubsection{Flight time}
The flight time describes the length of the swing phase while taking a stride. Using \ac{GRF}, the flight time is defined from the toe-off time to the following heel-strike of the same leg. As described in \cref{sec:flight_time_definition}, for the radar data, the flight time is defined from the time instant of maximal knee velocity to the end of the gait cycle.

\subsubsection{Step time}
The step time denotes the period of time taken for a step. Based on \ac{GRF}, the step time is defined from the heel-strike of one foot to the heel-strike of the following other foot. The step time is obtained by averaging the difference between two consecutive minima in the envelope signal. The averaged micro-Doppler signatures of full gait cycles are cut to two step signatures for the left and right foot based on the average step time of the first leg in the gait cycle.

\subsubsection{Cadence}
The cadence is typically defined as the number of steps per minute. For both, the \ac{GRF} and radar data, the cadence is calculated based on the reciprocal of the average step time (in min).

\subsubsection{Stride length}
The stride length describes the distance which is covered during one stride (or cycle). For both measurement systems, the stride length is calculated as the product of the stride time and the treadmill speed.

\subsubsection{Step length}
The step length describes the distance between successive heel-off events of the opposite feet. For both measurement systems, the step length is calculated as the product of the stride time and treadmill speed.

\subsubsection{Maximal foot velocity}
Using motion capture data, the radial velocity of the toe marker is averaged over all gait cycles. The maximum of the averaged velocities yields the maximal foot velocity (see \cref{moc_normal_fast_front,moc_abnormal_fast_front,moc_normal_slow_front,moc_normal_fast_back}). Given micro-Doppler signatures, the maximal foot velocity is determined by averaging the toe envelope signal over all gait cycles. The maximum of the averaged envelope signals yields the average maximal foot velocity (see \cref{rad_normal_fast_front,rad_abnormal_fast_front,rad_normal_slow_front,rad_normal_fast_back}). 

\subsubsection{Maximal ankle velocity}
Similarly to the extraction of the maximal foot velocity, the maximal ankle velocity is given by the maximum of the average radial velocity of the marker positioned at the ankle (see \cref{moc_normal_fast_front,moc_abnormal_fast_front,moc_normal_slow_front,moc_normal_fast_back}). In order to extract the average maximal ankle velocity from the radar data, the micro-Doppler envelope signal of the ankle is averaged over all gait cycles in a measurement. Its maximum gives the average maximal ankle velocity (see \cref{rad_normal_fast_front,rad_abnormal_fast_front,rad_normal_slow_front,rad_normal_fast_back}). 

\subsubsection{Maximal knee velocity}
The maximal knee velocity is extracted from the motion capture data by finding the maximal average velocity of the knee marker in the $y$-direction (see \cref{moc_normal_fast_front,moc_abnormal_fast_front,moc_normal_slow_front,moc_normal_fast_back}). Then, the knee envelope signal is found, as shown in \cref{fig:specs_thr}, and averaged over all gait cycles. Its maximum gives the average maximal knee velocity (see \cref{rad_normal_fast_front,rad_abnormal_fast_front,rad_normal_slow_front,rad_normal_fast_back}). In case of the knee envelope, the first local maximum is considered opposed to the global maximum (see e.g.~\cref{rad_env_abnormal_fast_front}). 

\subsubsection{Time instant of maximal knee velocity}
Having detected the maximal knee velocity in the motion capture and radar data, respectively, the corresponding time instants during the gait cycle serve as an additional characteristic to describe the degree of abnormality. As explained in \cref{sec:flight_time_definition}, for the radar data, we use the time instant of maximal knee velocity as the onset of the flight time.

\subsection{Hypothesis testing}

\label{sec:hypo_test}
To evaluate the accuracy of the radar system in measuring the aforementioned gait parameters in comparison to the motion capture system, we employ hypothesis testing. More specifically, we aim to show that there is no significant difference in measuring the gait parameters using radar compared to using motion capturing. For this, we resort to the Wilcoxon signed rank test, which is a non-parametric, paired, two-sided test for the null hypothesis that the differences come from a distribution with zero median \cite{Gib11}. The test is performed \textit{individually} for each gait parameter, knee angle restriction, viewing angle of the radar and walking speed under analysis, such that in total $N = 220$ tests are performed. Let $X_{r,i}$ and $X_{m,i}$ be \textit{one} of the gait parameters given in \cref{sec:gait_param_extr} obtained through radar and motion capturing, respectively, and $i=1,\dots,19$ refer to the nineteen test subjects. We consider the random sample of pairs $(X_{r,i},X_{m,i})$ and form the differences $D_i = X_{m,i} - X_{r,i}$. Then, we test the hypothesis
\begin{align*}
    \Hyp_0: \quad & M_D = 0 \,,
\end{align*}
where $M_D$ is the median of the population of differences $D_i$. Given observations $x_{m,i}$ and $x_{r,i}$, $i=1,\dots,19$, we form the differences $d_i = x_{m,i} - x_{r,i}$ and calculate the sum of signed ranks. Based on this test statistic, we calculate the $p$-value for a confidence level of $95$\% (confidence coefficient $\alpha = 0.05$). The $p$-value gives the probability of accepting the alternative hypothesis although the null hypothesis is true. Thus, in a first step, we test the hypotheses $\Hyp_{(1)}, \Hyp_{(2)},\dots,\Hyp_{(N)}$ by comparing the corresponding $p$-values to $\alpha$, where $\Hyp_{(j)}$ denotes the null hypothesis for the $j$th test and $N = 220$. For $p \le 0.05$, we reject the null hypothesis.

In a second step, we control the expected proportion of falsely rejected hypotheses by applying the Benjamini-Hochberg (BH) procedure which controls the false discovery rate (FDR) \cite{Ben95}. For this, rank ordering of the calculated $p$-values is applied, such that $p_{(1)} \leq p_{(2)} \leq \cdots \leq p_{(N)}$. Then, we find $k = \{ \max j~|~p_{(j)} \leq \frac{j}{N} \alpha \}$. Thus, through the BH procedure, we obtain $p^\ast \coloneqq p_{(k)}$, which serves as a significance threshold for the $p$-values, i.e., for $p \leq p^\ast$ the null hypothesis is rejected ($\tikzcircle[black, fill=myred!70]{1.5pt}$). For $p > p^\ast$ we accept the null hypothesis ($\tikzcircle[black, fill=mygreen!70]{2.2pt}$). 

Since very large $p$-values generally indicate high confidence in the null hypothesis, we highlight $p$-values with $p > 0.8$ ($\tikzcircle[black, fill=myturkis!70]{3pt}$).

\section{Results}
\begin{table*}[t!]
\centering
\setlength{\tabcolsep}{2.5pt} 
\newlength{\lengtha} \setlength{\lengtha}{3.5mm}
\renewcommand{\arraystretch}{1.2}
\caption{Results of testing the null hypothesis that the median of the differences in the parameter measured by the motion-capture and radar system is zero. For each treadmill speed ($0.7$\,m/s, $1.1$\,m/s) and radar position (front, back), the results are shown for increasing knee restriction from left to right (no, $45$\textdegree, $30$\textdegree, $20$\textdegree, $10$\textdegree). The red dot symbol ($\tikzcircle[black, fill=myred!70]{1.5pt}$) indicates that the median is statistically significantly different from zero ($p < 0.0057$). The green and blue circles ($\tikzcircle[black, fill=mygreen!70]{2.2pt}$) and ($\tikzcircle[black, fill=myturkis!70]{3pt}$) indicate that the null hypothesis cannot be rejected with $p > 0.0057$ and $p > 0.8$, respectively.}
\label{tab:results}
\begin{tabular}{r l@{\hskip \lengtha} *{4}{*{5}{c}@{\extracolsep{\fill}\hskip \lengtha}}}\toprule
 & & \multicolumn{10}{c@{\hskip \lengtha}}{\textbf{0.7\,m/s}} &            \multicolumn{10}{c@{\hskip \lengtha}}{\textbf{1.1\,m/s}}\\
 \cmidrule(l{-0.1em}r{1em}){3-12} \cmidrule(l{-0.1em}r{1em}){13-22}
& & \multicolumn{5}{c@{\hskip \lengtha}}{front} & \multicolumn{5}{c@{\hskip \lengtha}}{back}
  & \multicolumn{5}{c@{\hskip \lengtha}}{front} & \multicolumn{5}{c@{\hskip \lengtha}}{back}\\\midrule[\heavyrulewidth]
1) & stride time& 	$\tikzcircle[black, fill=mygreen!70]{2.2pt}$& 	$\tikzcircle[black, fill=mygreen!70]{2.2pt}$& 	$\tikzcircle[black, fill=mygreen!70]{2.2pt}$& 	$\tikzcircle[black, fill=mygreen!70]{2.2pt}$& 	$\tikzcircle[black, fill=mygreen!70]{2.2pt}$& 	$\tikzcircle[black, fill=mygreen!70]{2.2pt}$& 	$\tikzcircle[black, fill=mygreen!70]{2.2pt}$& 	$\tikzcircle[black, fill=mygreen!70]{2.2pt}$& 	$\tikzcircle[black, fill=mygreen!70]{2.2pt}$& $\tikzcircle[black, fill=myturkis!70]{3pt}$	& 	$\tikzcircle[black, fill=mygreen!70]{2.2pt}$& 	$\tikzcircle[black, fill=mygreen!70]{2.2pt}$& 	$\tikzcircle[black, fill=mygreen!70]{2.2pt}$& 	$\tikzcircle[black, fill=mygreen!70]{2.2pt}$& 	$\tikzcircle[black, fill=mygreen!70]{2.2pt}$& 	$\tikzcircle[black, fill=mygreen!70]{2.2pt}$& 	$\tikzcircle[black, fill=mygreen!70]{2.2pt}$& 	$\tikzcircle[black, fill=mygreen!70]{2.2pt}$& 	$\tikzcircle[black, fill=mygreen!70]{2.2pt}$& $\tikzcircle[black, fill=myturkis!70]{3pt}$		\\
2) & stance time& $\tikzcircle[black, fill=myturkis!70]{3pt}$	& 	$\tikzcircle[black, fill=mygreen!70]{2.2pt}$& 	$\tikzcircle[black, fill=mygreen!70]{2.2pt}$& 	$\tikzcircle[black, fill=mygreen!70]{2.2pt}$& 	$\tikzcircle[black, fill=mygreen!70]{2.2pt}$& 	$\tikzcircle[black, fill=mygreen!70]{2.2pt}$& 	$\tikzcircle[black, fill=mygreen!70]{2.2pt}$& 	$\tikzcircle[black, fill=mygreen!70]{2.2pt}$& 	$\tikzcircle[black, fill=mygreen!70]{2.2pt}$& $\tikzcircle[black, fill=myturkis!70]{3pt}$	& $\tikzcircle[black, fill=myred!70]{1.5pt}$	& $\tikzcircle[black, fill=myred!70]{1.5pt}$	& $\tikzcircle[black, fill=myred!70]{1.5pt}$	& $\tikzcircle[black, fill=myred!70]{1.5pt}$	& 	$\tikzcircle[black, fill=mygreen!70]{2.2pt}$& 	$\tikzcircle[black, fill=mygreen!70]{2.2pt}$& 	$\tikzcircle[black, fill=mygreen!70]{2.2pt}$& 	$\tikzcircle[black, fill=mygreen!70]{2.2pt}$& 	$\tikzcircle[black, fill=mygreen!70]{2.2pt}$& 	$\tikzcircle[black, fill=mygreen!70]{2.2pt}$	\\
3) & flight time& $\tikzcircle[black, fill=myturkis!70]{3pt}$	& 	$\tikzcircle[black, fill=mygreen!70]{2.2pt}$& 	$\tikzcircle[black, fill=mygreen!70]{2.2pt}$& 	$\tikzcircle[black, fill=mygreen!70]{2.2pt}$& 	$\tikzcircle[black, fill=mygreen!70]{2.2pt}$& 	$\tikzcircle[black, fill=mygreen!70]{2.2pt}$& 	$\tikzcircle[black, fill=mygreen!70]{2.2pt}$& 	$\tikzcircle[black, fill=mygreen!70]{2.2pt}$& 	$\tikzcircle[black, fill=mygreen!70]{2.2pt}$& $\tikzcircle[black, fill=myturkis!70]{3pt}$	& $\tikzcircle[black, fill=myred!70]{1.5pt}$	& $\tikzcircle[black, fill=myred!70]{1.5pt}$	& $\tikzcircle[black, fill=myred!70]{1.5pt}$	& $\tikzcircle[black, fill=myred!70]{1.5pt}$	& 	$\tikzcircle[black, fill=mygreen!70]{2.2pt}$& 	$\tikzcircle[black, fill=mygreen!70]{2.2pt}$& 	$\tikzcircle[black, fill=mygreen!70]{2.2pt}$& 	$\tikzcircle[black, fill=mygreen!70]{2.2pt}$& 	$\tikzcircle[black, fill=mygreen!70]{2.2pt}$& 	$\tikzcircle[black, fill=mygreen!70]{2.2pt}$	\\
4) & step time& 	$\tikzcircle[black, fill=mygreen!70]{2.2pt}$& $\tikzcircle[black, fill=myturkis!70]{3pt}$	& 	$\tikzcircle[black, fill=mygreen!70]{2.2pt}$& 	$\tikzcircle[black, fill=mygreen!70]{2.2pt}$& 	$\tikzcircle[black, fill=mygreen!70]{2.2pt}$& 	$\tikzcircle[black, fill=mygreen!70]{2.2pt}$& 	$\tikzcircle[black, fill=mygreen!70]{2.2pt}$& 	$\tikzcircle[black, fill=mygreen!70]{2.2pt}$& 	$\tikzcircle[black, fill=mygreen!70]{2.2pt}$& 	$\tikzcircle[black, fill=mygreen!70]{2.2pt}$& 	$\tikzcircle[black, fill=mygreen!70]{2.2pt}$& 	$\tikzcircle[black, fill=mygreen!70]{2.2pt}$& 	$\tikzcircle[black, fill=mygreen!70]{2.2pt}$& 	$\tikzcircle[black, fill=mygreen!70]{2.2pt}$& 	$\tikzcircle[black, fill=mygreen!70]{2.2pt}$& $\tikzcircle[black, fill=myturkis!70]{3pt}$	& 	$\tikzcircle[black, fill=mygreen!70]{2.2pt}$& 	$\tikzcircle[black, fill=mygreen!70]{2.2pt}$& 	$\tikzcircle[black, fill=mygreen!70]{2.2pt}$& 	$\tikzcircle[black, fill=mygreen!70]{2.2pt}$	\\
5) & cadence& $\tikzcircle[black, fill=myturkis!70]{3pt}$	& 	$\tikzcircle[black, fill=mygreen!70]{2.2pt}$& 	$\tikzcircle[black, fill=mygreen!70]{2.2pt}$& 	$\tikzcircle[black, fill=mygreen!70]{2.2pt}$& 	$\tikzcircle[black, fill=mygreen!70]{2.2pt}$& 	$\tikzcircle[black, fill=mygreen!70]{2.2pt}$& 	$\tikzcircle[black, fill=mygreen!70]{2.2pt}$& 	$\tikzcircle[black, fill=mygreen!70]{2.2pt}$& 	$\tikzcircle[black, fill=mygreen!70]{2.2pt}$& 	$\tikzcircle[black, fill=mygreen!70]{2.2pt}$& 	$\tikzcircle[black, fill=mygreen!70]{2.2pt}$& 	$\tikzcircle[black, fill=mygreen!70]{2.2pt}$& $\tikzcircle[black, fill=myturkis!70]{3pt}$	& 	$\tikzcircle[black, fill=mygreen!70]{2.2pt}$& 	$\tikzcircle[black, fill=mygreen!70]{2.2pt}$& 	$\tikzcircle[black, fill=mygreen!70]{2.2pt}$& 	$\tikzcircle[black, fill=mygreen!70]{2.2pt}$& 	$\tikzcircle[black, fill=mygreen!70]{2.2pt}$& 	$\tikzcircle[black, fill=mygreen!70]{2.2pt}$& 	$\tikzcircle[black, fill=mygreen!70]{2.2pt}$	\\
6) & stride length& 	$\tikzcircle[black, fill=mygreen!70]{2.2pt}$& 	$\tikzcircle[black, fill=mygreen!70]{2.2pt}$& 	$\tikzcircle[black, fill=mygreen!70]{2.2pt}$& 	$\tikzcircle[black, fill=mygreen!70]{2.2pt}$& 	$\tikzcircle[black, fill=mygreen!70]{2.2pt}$& 	$\tikzcircle[black, fill=mygreen!70]{2.2pt}$& 	$\tikzcircle[black, fill=mygreen!70]{2.2pt}$& 	$\tikzcircle[black, fill=mygreen!70]{2.2pt}$& 	$\tikzcircle[black, fill=mygreen!70]{2.2pt}$& $\tikzcircle[black, fill=myturkis!70]{3pt}$	& 	$\tikzcircle[black, fill=mygreen!70]{2.2pt}$& 	$\tikzcircle[black, fill=mygreen!70]{2.2pt}$& 	$\tikzcircle[black, fill=mygreen!70]{2.2pt}$& 	$\tikzcircle[black, fill=mygreen!70]{2.2pt}$& 	$\tikzcircle[black, fill=mygreen!70]{2.2pt}$& 	$\tikzcircle[black, fill=mygreen!70]{2.2pt}$& 	$\tikzcircle[black, fill=mygreen!70]{2.2pt}$& 	$\tikzcircle[black, fill=mygreen!70]{2.2pt}$& 	$\tikzcircle[black, fill=mygreen!70]{2.2pt}$& $\tikzcircle[black, fill=myturkis!70]{3pt}$		\\
7) & step length& 	$\tikzcircle[black, fill=mygreen!70]{2.2pt}$& $\tikzcircle[black, fill=myturkis!70]{3pt}$	& 	$\tikzcircle[black, fill=mygreen!70]{2.2pt}$& 	$\tikzcircle[black, fill=mygreen!70]{2.2pt}$& 	$\tikzcircle[black, fill=mygreen!70]{2.2pt}$& 	$\tikzcircle[black, fill=mygreen!70]{2.2pt}$& 	$\tikzcircle[black, fill=mygreen!70]{2.2pt}$& 	$\tikzcircle[black, fill=mygreen!70]{2.2pt}$& 	$\tikzcircle[black, fill=mygreen!70]{2.2pt}$& 	$\tikzcircle[black, fill=mygreen!70]{2.2pt}$& 	$\tikzcircle[black, fill=mygreen!70]{2.2pt}$& 	$\tikzcircle[black, fill=mygreen!70]{2.2pt}$& 	$\tikzcircle[black, fill=mygreen!70]{2.2pt}$& 	$\tikzcircle[black, fill=mygreen!70]{2.2pt}$& 	$\tikzcircle[black, fill=mygreen!70]{2.2pt}$& $\tikzcircle[black, fill=myturkis!70]{3pt}$	& 	$\tikzcircle[black, fill=mygreen!70]{2.2pt}$& 	$\tikzcircle[black, fill=mygreen!70]{2.2pt}$& 	$\tikzcircle[black, fill=mygreen!70]{2.2pt}$& 	$\tikzcircle[black, fill=mygreen!70]{2.2pt}$	\\
8) & maximal foot velocity& $\tikzcircle[black, fill=myturkis!70]{3pt}$	& 	$\tikzcircle[black, fill=mygreen!70]{2.2pt}$& 	$\tikzcircle[black, fill=mygreen!70]{2.2pt}$& 	$\tikzcircle[black, fill=mygreen!70]{2.2pt}$& $\tikzcircle[black, fill=myturkis!70]{3pt}$	& 	$\tikzcircle[black, fill=mygreen!70]{2.2pt}$& 	$\tikzcircle[black, fill=mygreen!70]{2.2pt}$& $\tikzcircle[black, fill=myred!70]{1.5pt}$	& 	$\tikzcircle[black, fill=mygreen!70]{2.2pt}$& $\tikzcircle[black, fill=myred!70]{1.5pt}$	& 	$\tikzcircle[black, fill=mygreen!70]{2.2pt}$& 	$\tikzcircle[black, fill=mygreen!70]{2.2pt}$& 	$\tikzcircle[black, fill=mygreen!70]{2.2pt}$& 	$\tikzcircle[black, fill=mygreen!70]{2.2pt}$& 	$\tikzcircle[black, fill=mygreen!70]{2.2pt}$& 	$\tikzcircle[black, fill=mygreen!70]{2.2pt}$& 	$\tikzcircle[black, fill=mygreen!70]{2.2pt}$& 	$\tikzcircle[black, fill=mygreen!70]{2.2pt}$& $\tikzcircle[black, fill=myturkis!70]{3pt}$	& $\tikzcircle[black, fill=myturkis!70]{3pt}$		\\
9) & maximal ankle velocity& 	$\tikzcircle[black, fill=mygreen!70]{2.2pt}$& 	$\tikzcircle[black, fill=mygreen!70]{2.2pt}$& 	$\tikzcircle[black, fill=mygreen!70]{2.2pt}$& 	$\tikzcircle[black, fill=mygreen!70]{2.2pt}$& 	$\tikzcircle[black, fill=mygreen!70]{2.2pt}$& $\tikzcircle[black, fill=myred!70]{1.5pt}$	& 	$\tikzcircle[black, fill=mygreen!70]{2.2pt}$& $\tikzcircle[black, fill=myred!70]{1.5pt}$	& 	$\tikzcircle[black, fill=mygreen!70]{2.2pt}$& 	$\tikzcircle[black, fill=mygreen!70]{2.2pt}$& $\tikzcircle[black, fill=myturkis!70]{3pt}$	& 	$\tikzcircle[black, fill=mygreen!70]{2.2pt}$& 	$\tikzcircle[black, fill=mygreen!70]{2.2pt}$& 	$\tikzcircle[black, fill=mygreen!70]{2.2pt}$& 	$\tikzcircle[black, fill=mygreen!70]{2.2pt}$& 	$\tikzcircle[black, fill=mygreen!70]{2.2pt}$& $\tikzcircle[black, fill=myturkis!70]{3pt}$	& 	$\tikzcircle[black, fill=mygreen!70]{2.2pt}$& $\tikzcircle[black, fill=myturkis!70]{3pt}$	& $\tikzcircle[black, fill=myturkis!70]{3pt}$		\\
10) & maximal knee velocity& 	$\tikzcircle[black, fill=mygreen!70]{2.2pt}$& $\tikzcircle[black, fill=myred!70]{1.5pt}$	& 	$\tikzcircle[black, fill=mygreen!70]{2.2pt}$& 	$\tikzcircle[black, fill=mygreen!70]{2.2pt}$& $\tikzcircle[black, fill=myred!70]{1.5pt}$	& $\tikzcircle[black, fill=myred!70]{1.5pt}$	& 	$\tikzcircle[black, fill=mygreen!70]{2.2pt}$& 	$\tikzcircle[black, fill=mygreen!70]{2.2pt}$& 	$\tikzcircle[black, fill=mygreen!70]{2.2pt}$& $\tikzcircle[black, fill=myturkis!70]{3pt}$	& $\tikzcircle[black, fill=myred!70]{1.5pt}$	& 	$\tikzcircle[black, fill=mygreen!70]{2.2pt}$& 	$\tikzcircle[black, fill=mygreen!70]{2.2pt}$& $\tikzcircle[black, fill=myturkis!70]{3pt}$	& 	$\tikzcircle[black, fill=mygreen!70]{2.2pt}$& $\tikzcircle[black, fill=myred!70]{1.5pt}$	& 	$\tikzcircle[black, fill=mygreen!70]{2.2pt}$& 	$\tikzcircle[black, fill=mygreen!70]{2.2pt}$& 	$\tikzcircle[black, fill=mygreen!70]{2.2pt}$& 	$\tikzcircle[black, fill=mygreen!70]{2.2pt}$	\\
11) & time instant of maximal knee velocity& $\tikzcircle[black, fill=myred!70]{1.5pt}$	& $\tikzcircle[black, fill=myred!70]{1.5pt}$	& 	$\tikzcircle[black, fill=mygreen!70]{2.2pt}$& 	$\tikzcircle[black, fill=mygreen!70]{2.2pt}$& $\tikzcircle[black, fill=myred!70]{1.5pt}$	& 	$\tikzcircle[black, fill=mygreen!70]{2.2pt}$& 	$\tikzcircle[black, fill=mygreen!70]{2.2pt}$& 	$\tikzcircle[black, fill=mygreen!70]{2.2pt}$& 	$\tikzcircle[black, fill=mygreen!70]{2.2pt}$& $\tikzcircle[black, fill=myturkis!70]{3pt}$	& $\tikzcircle[black, fill=myred!70]{1.5pt}$	& $\tikzcircle[black, fill=myred!70]{1.5pt}$	& $\tikzcircle[black, fill=myred!70]{1.5pt}$	& $\tikzcircle[black, fill=myred!70]{1.5pt}$	& 	$\tikzcircle[black, fill=mygreen!70]{2.2pt}$& $\tikzcircle[black, fill=myred!70]{1.5pt}$	& 	$\tikzcircle[black, fill=mygreen!70]{2.2pt}$& 	$\tikzcircle[black, fill=mygreen!70]{2.2pt}$& 	$\tikzcircle[black, fill=mygreen!70]{2.2pt}$& 	$\tikzcircle[black, fill=mygreen!70]{2.2pt}$	\\
\bottomrule
\end{tabular}
\end{table*}

\cref{tab:results} concisely presents the results of the applied Wilcoxon signed-rank test and subsequent BH procedure. The latter yields $p^\ast = 0.0057$ for the available data. From \cref{tab:results}, it can be seen that the radar-based estimates for the stride time, step time, cadence, stride length and step length generally agree with the parameters obtained from motion capture data, i.e., the respective differences are small. This holds for both treadmill speeds as well as for both radar positions. The stance and flight time are accurately measured by the radar system for all conditions at the slow treadmill speed ($0.7$\,m/s). At $1.1$\,m/s and when the radar has a back view on the target, both parameters can reliably be detected for all conditions except for unconstrained gait. However, both parameters are difficult to be extracted from the radar data at $1.1$\,m/s when the radar is positioned in front of the subject. 
The maximal toe, ankle and knee velocities can generally be extracted accurately from the radar data. Toe and ankle velocities are not correctly extracted in two out of the $20$ scenarios, knee velocities in five. 
The time index of maximal knee velocity cannot be reliably detected using radar when the radar is positioned in front of the subject. However, when the radar is positioned behind the subject, the parameter can be reliably obtained from the radar data, except for the case of unconstrained gait at $1.1$\,m/s.

\begin{figure*}[p]
  	\centering{
	    \subfloat[stride time]{\includegraphics[clip,trim= 0 0 20 0,width=0.68\columnwidth]{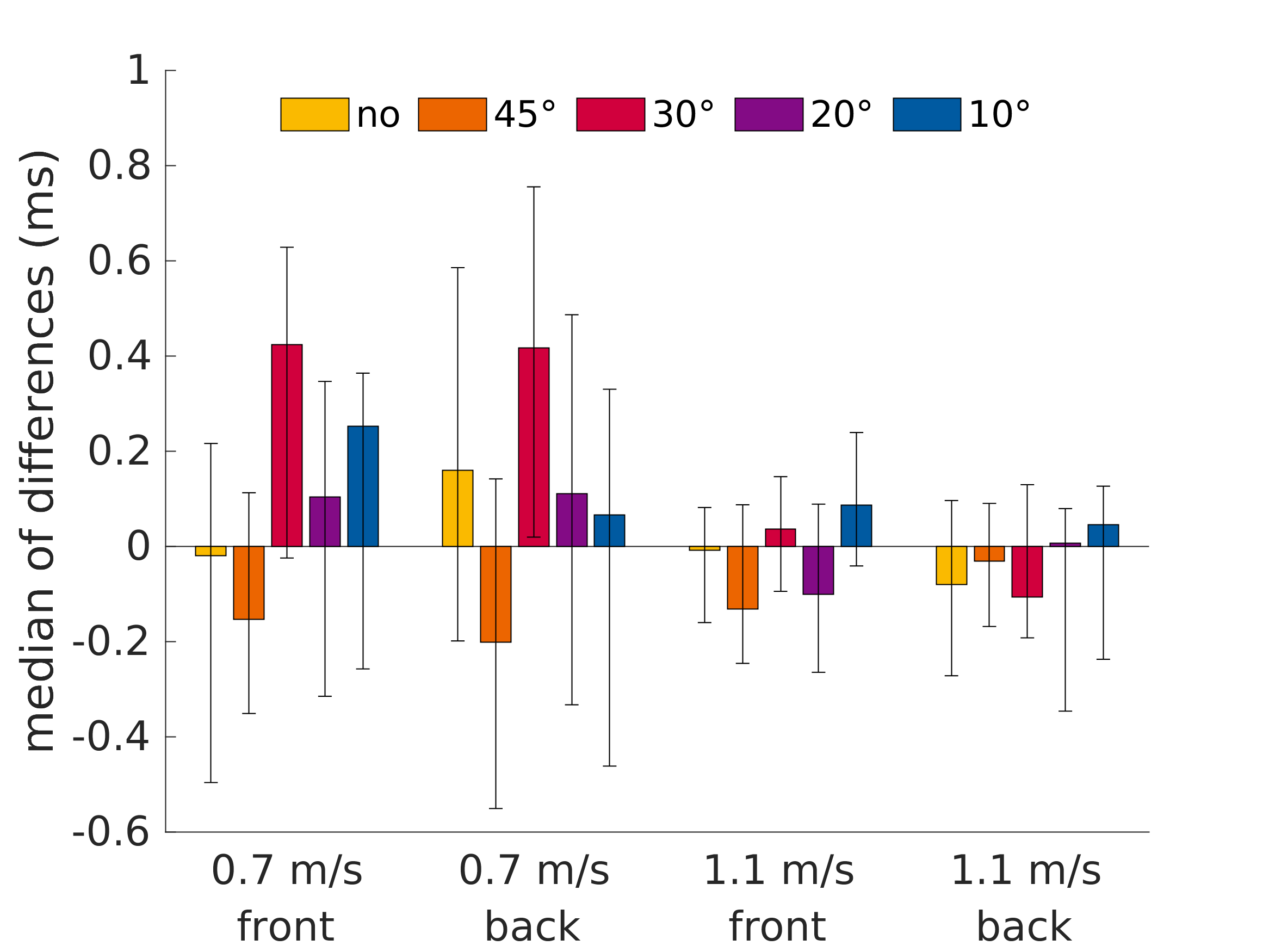}%
			\label{subfig:stridetime}}\hfill
		\subfloat[stance time]{\includegraphics[clip,trim= 0 0 20 10, width=0.68\columnwidth]{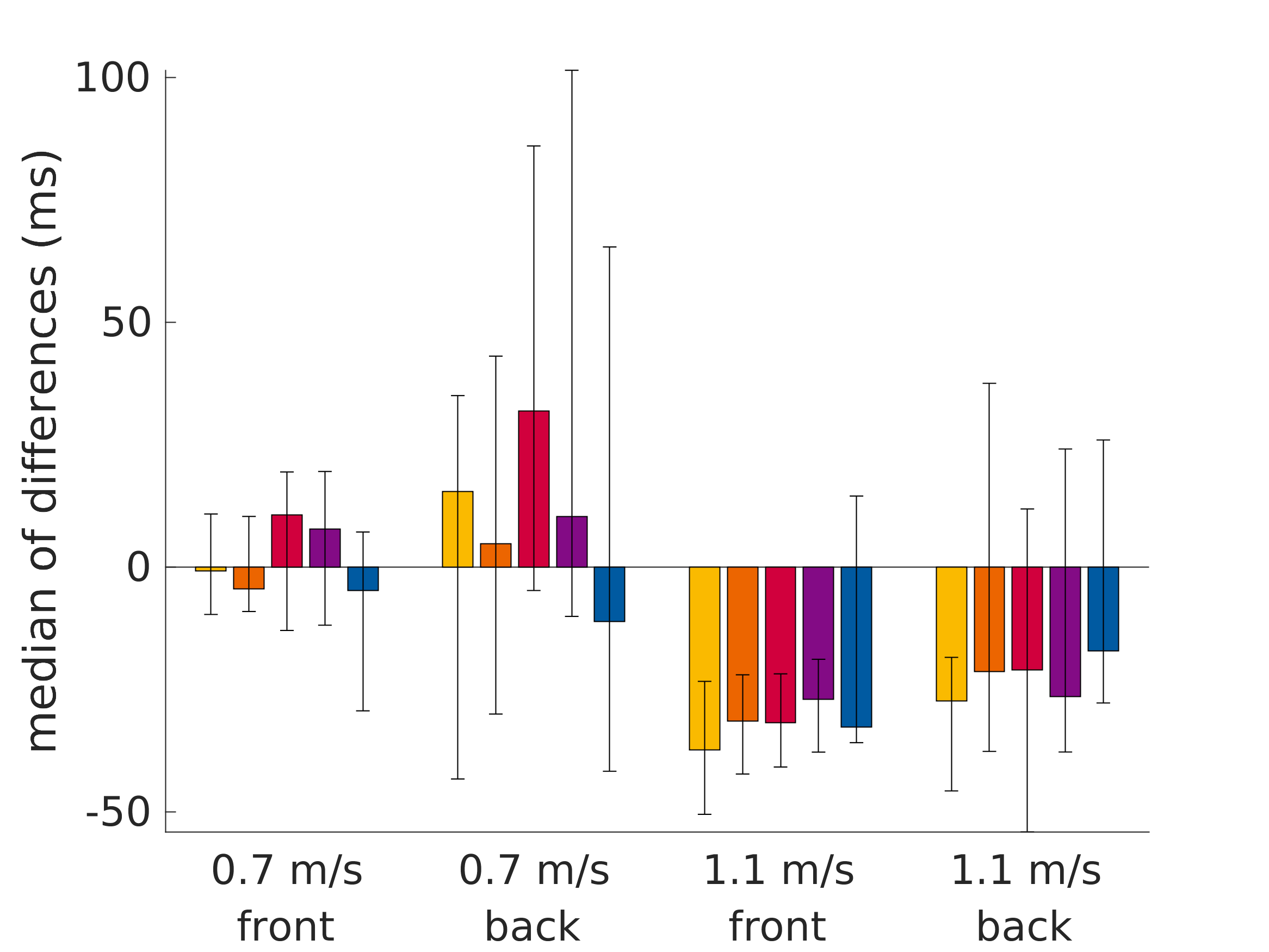}%
			\label{subfig:stancetime}}\hfill
  		\subfloat[flight time]{\includegraphics[clip,trim= 0 0 20 10,width=0.68\columnwidth]{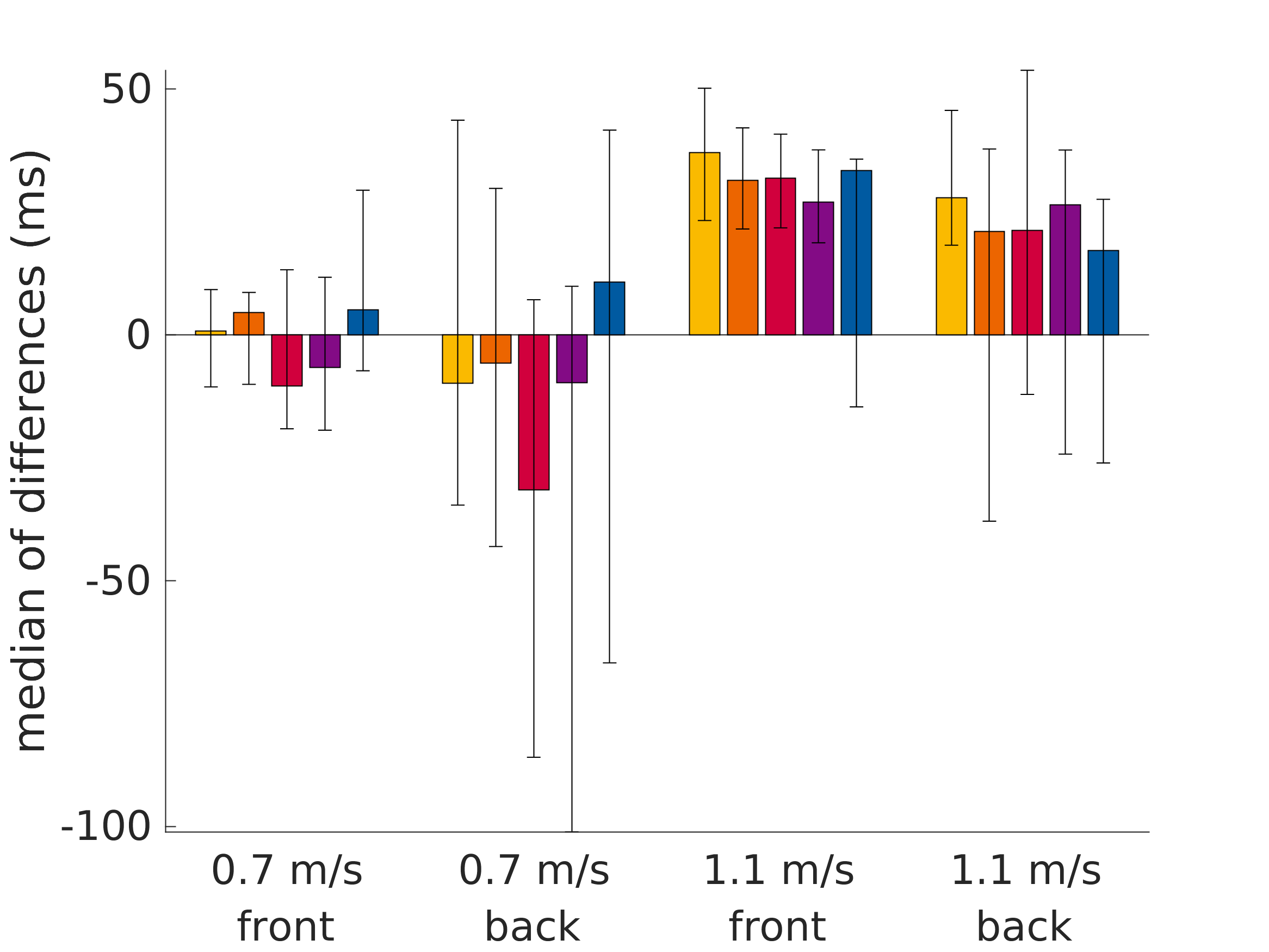}%
			\label{subfig:flighttime}}\\ \vspace{-1em}
  	    \subfloat[step time]{\includegraphics[clip,trim= 0 0 20 0,width=0.68\columnwidth]{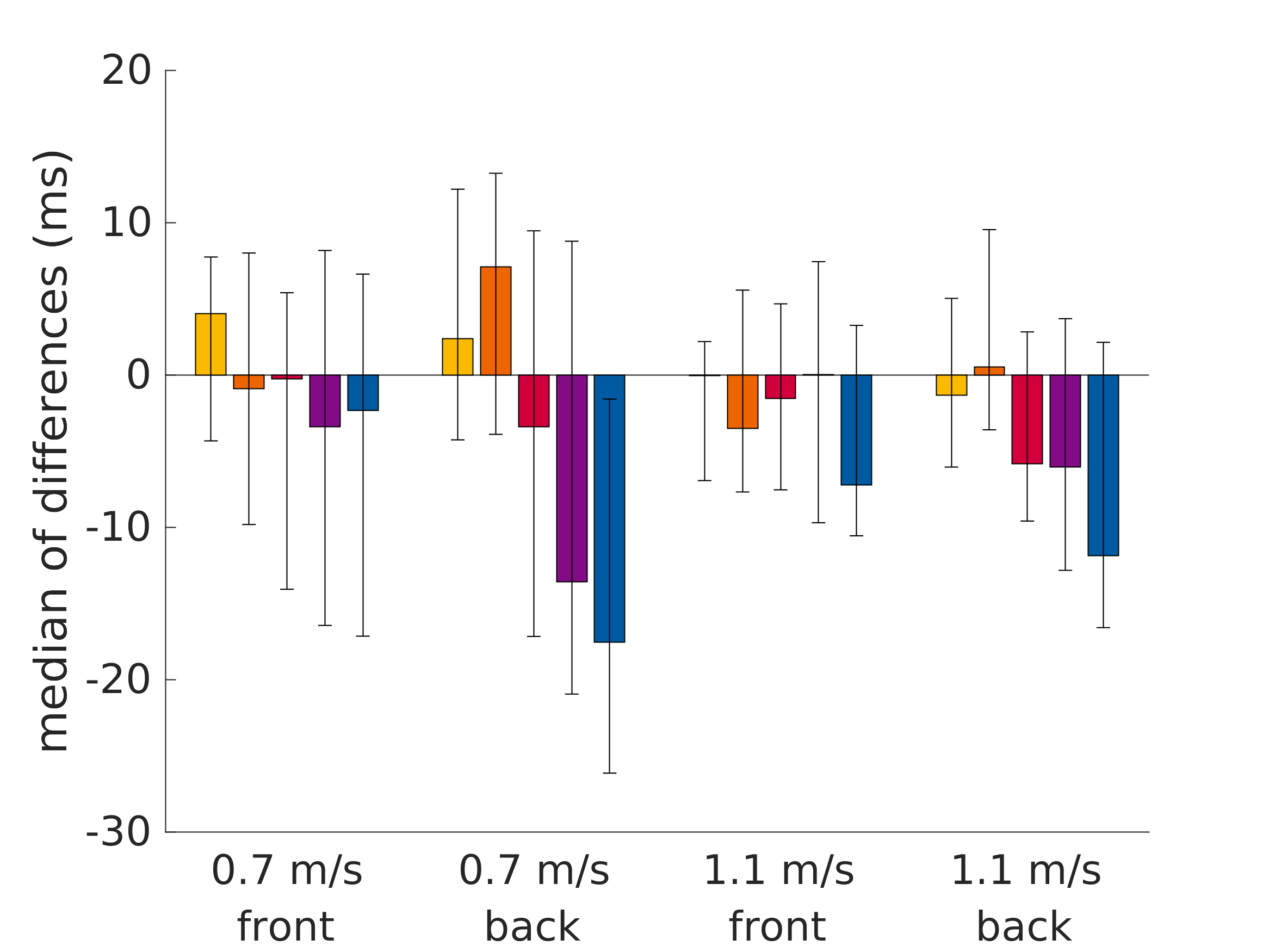}%
			\label{subfig:steptime}}\hfill
		\subfloat[cadence]{\includegraphics[clip,trim= 0 0 20 10, width=0.68\columnwidth]{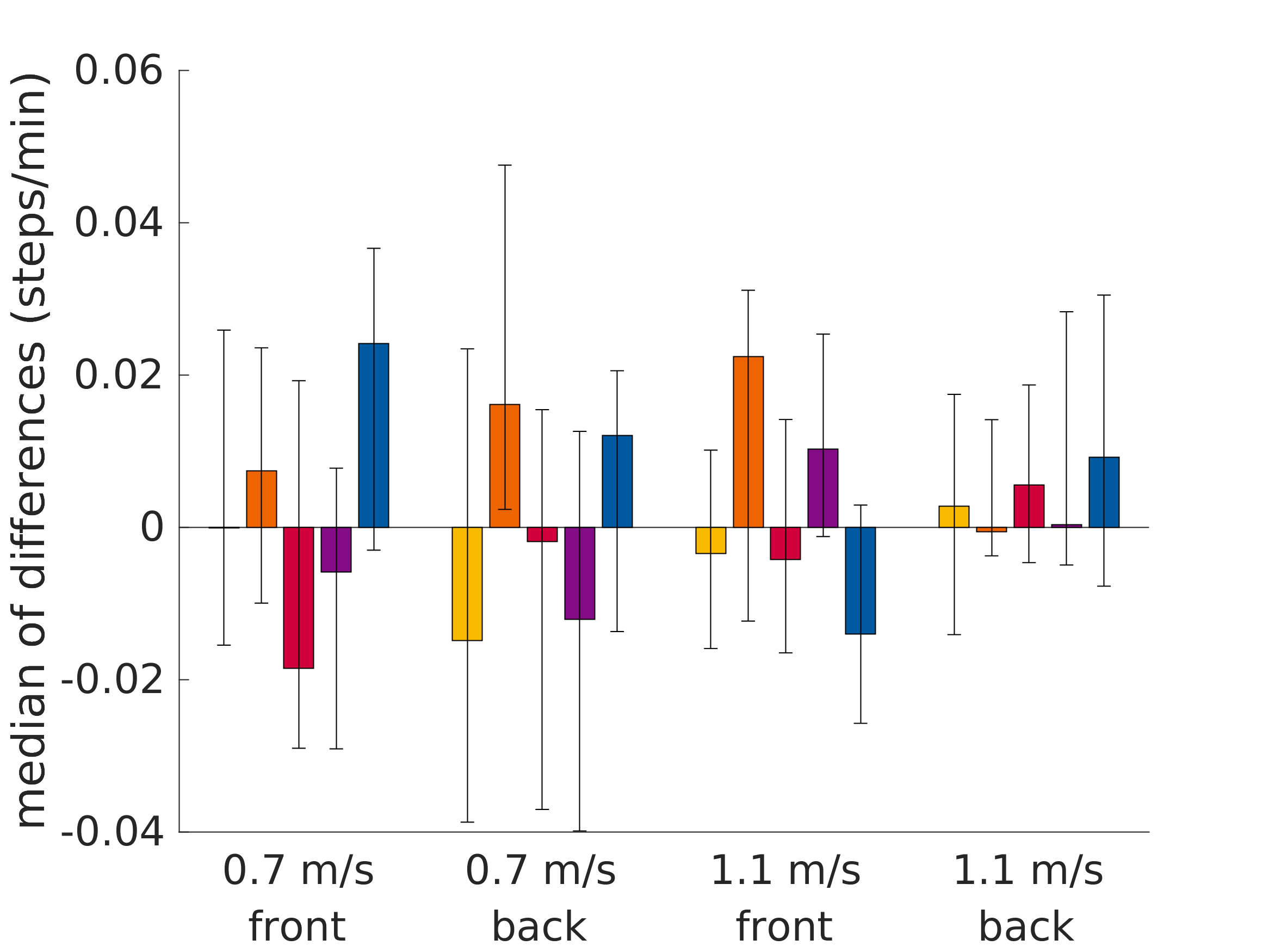}%
			\label{subfig:cadence}}\hfill
  		\subfloat[stride length]{\includegraphics[clip,trim= 0 0 20 10,width=0.68\columnwidth]{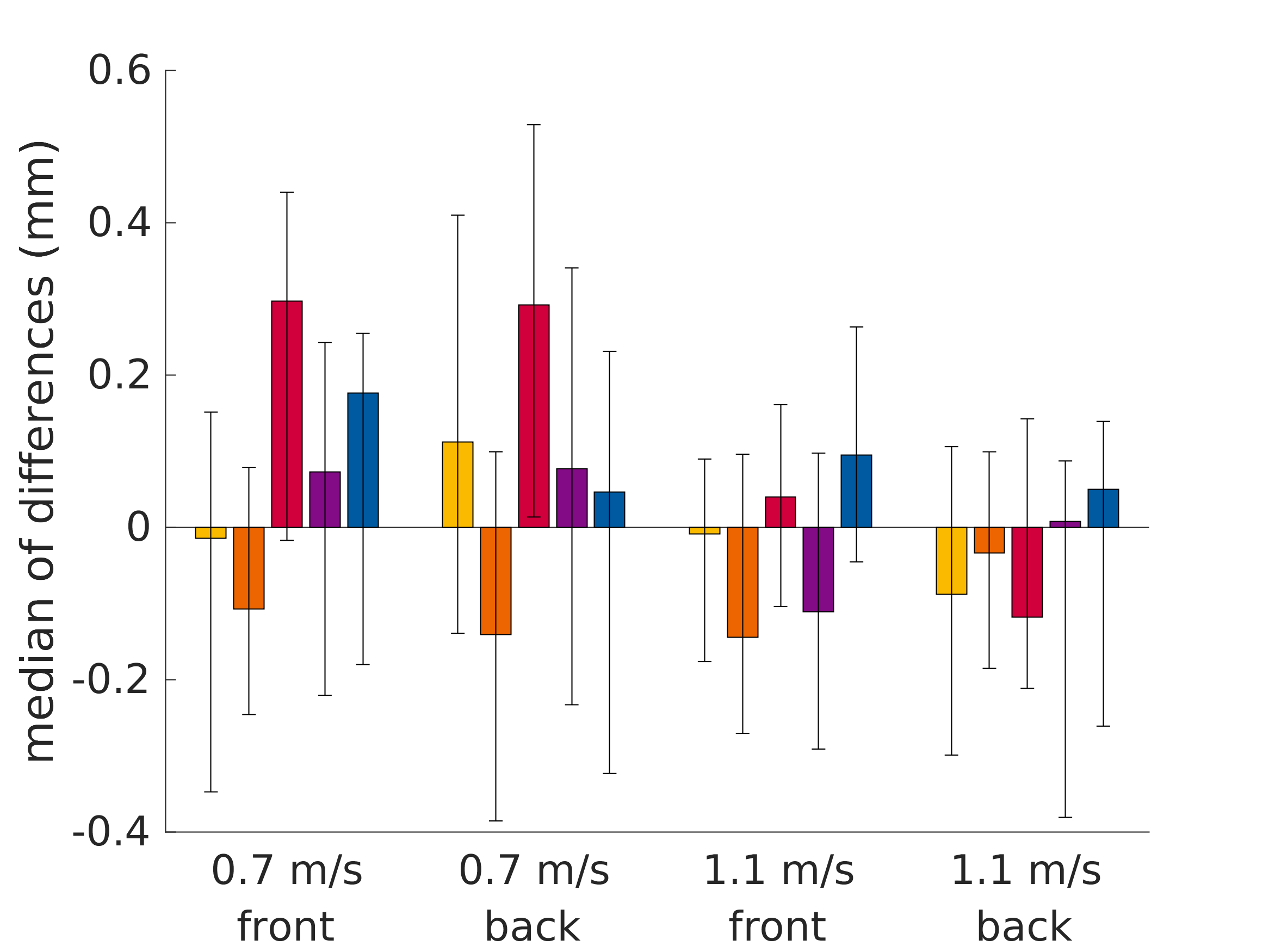}%
			\label{subfig:stridelength}}\\ \vspace{-1em}
		\subfloat[step length]{\includegraphics[clip,trim= 0 0 20 10,width=0.68\columnwidth]{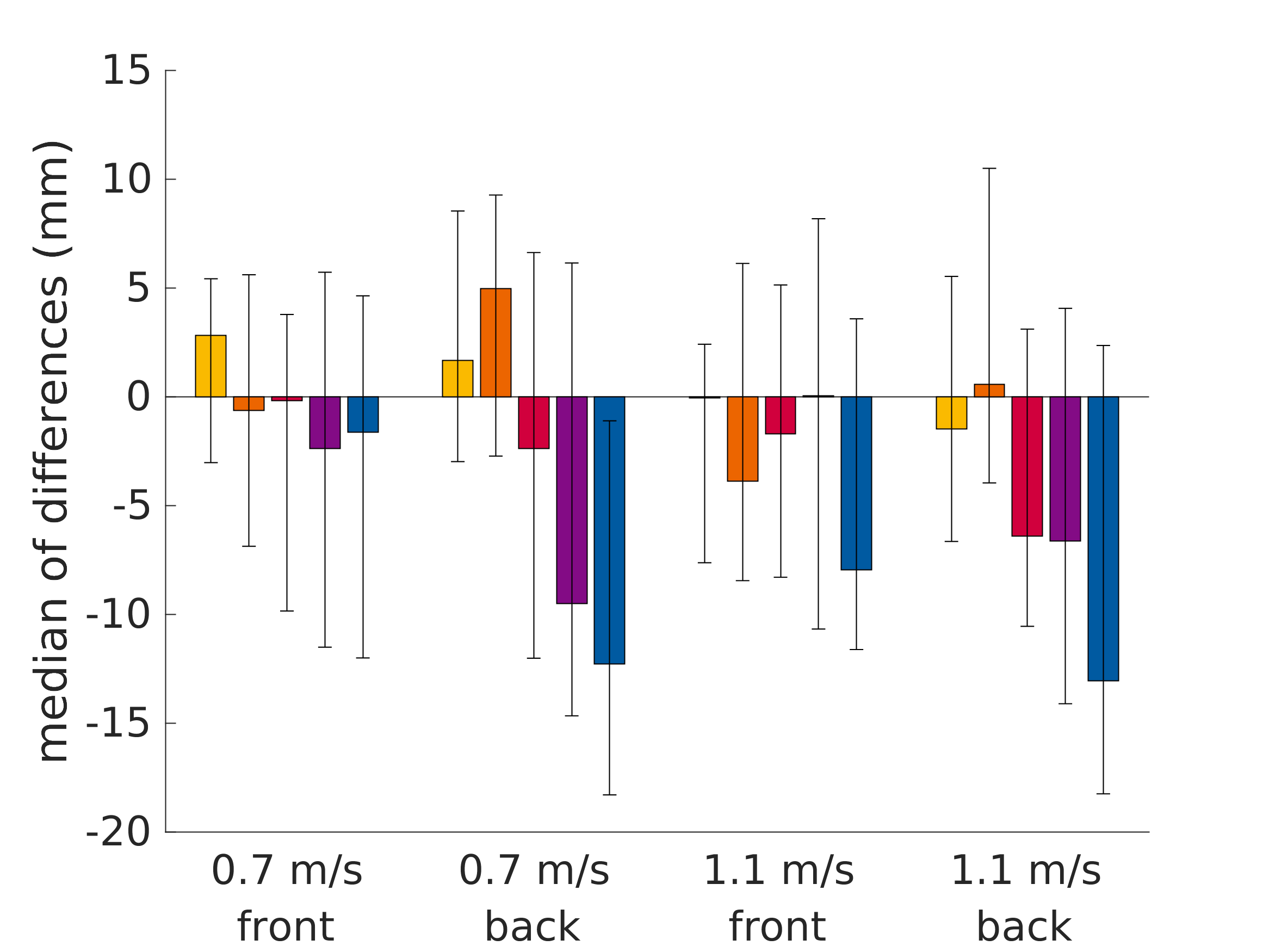}%
			\label{subfig:steplength}}\hfill
	    \subfloat[maximal foot velocity]{\includegraphics[clip,trim= 0 0 20 0,width=0.68\columnwidth]{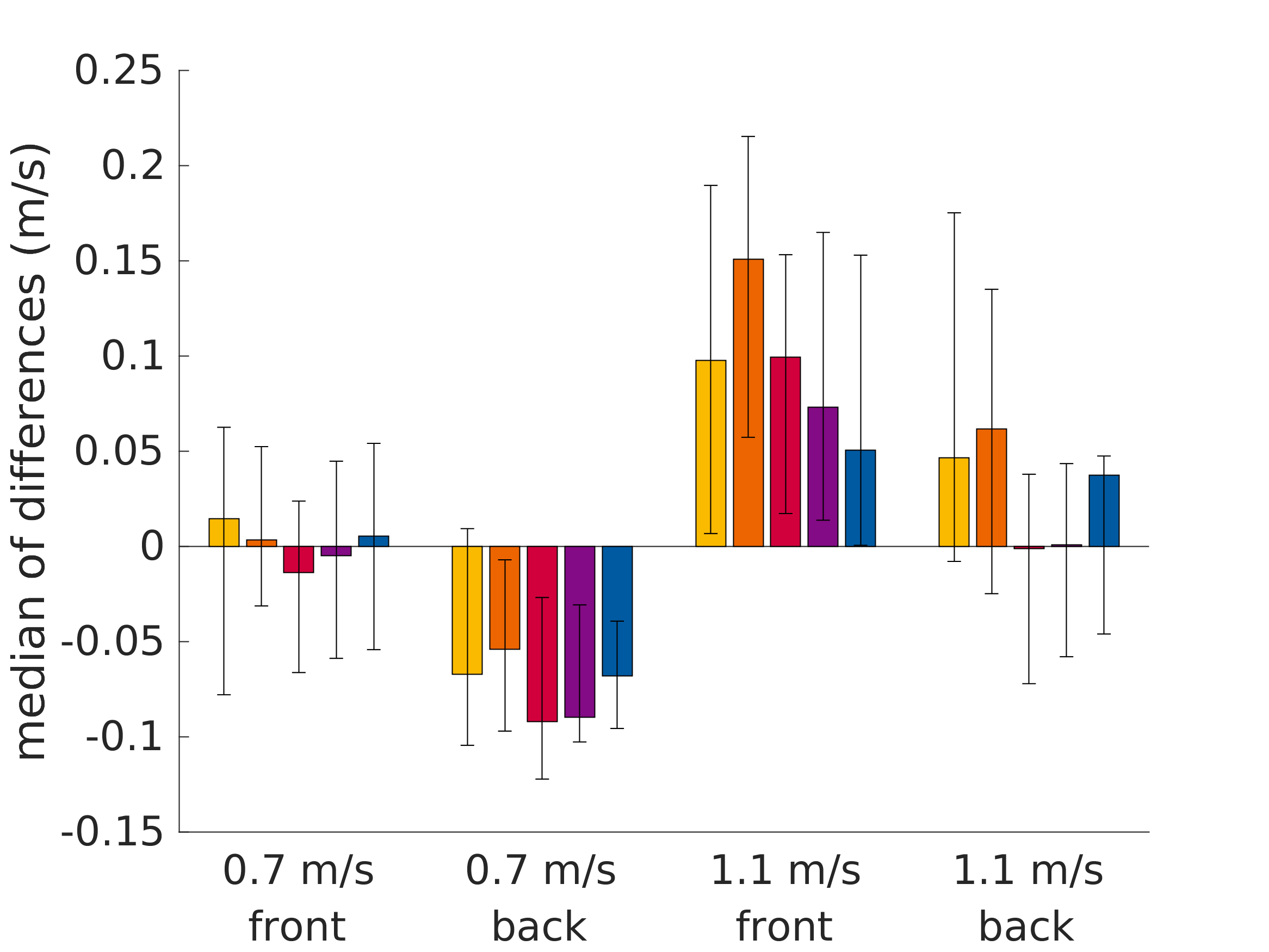}%
			\label{subfig:maxfoot}}\hfill
		\subfloat[maximal ankle velocity]{\includegraphics[clip,trim= 0 0 20 10, width=0.68\columnwidth]{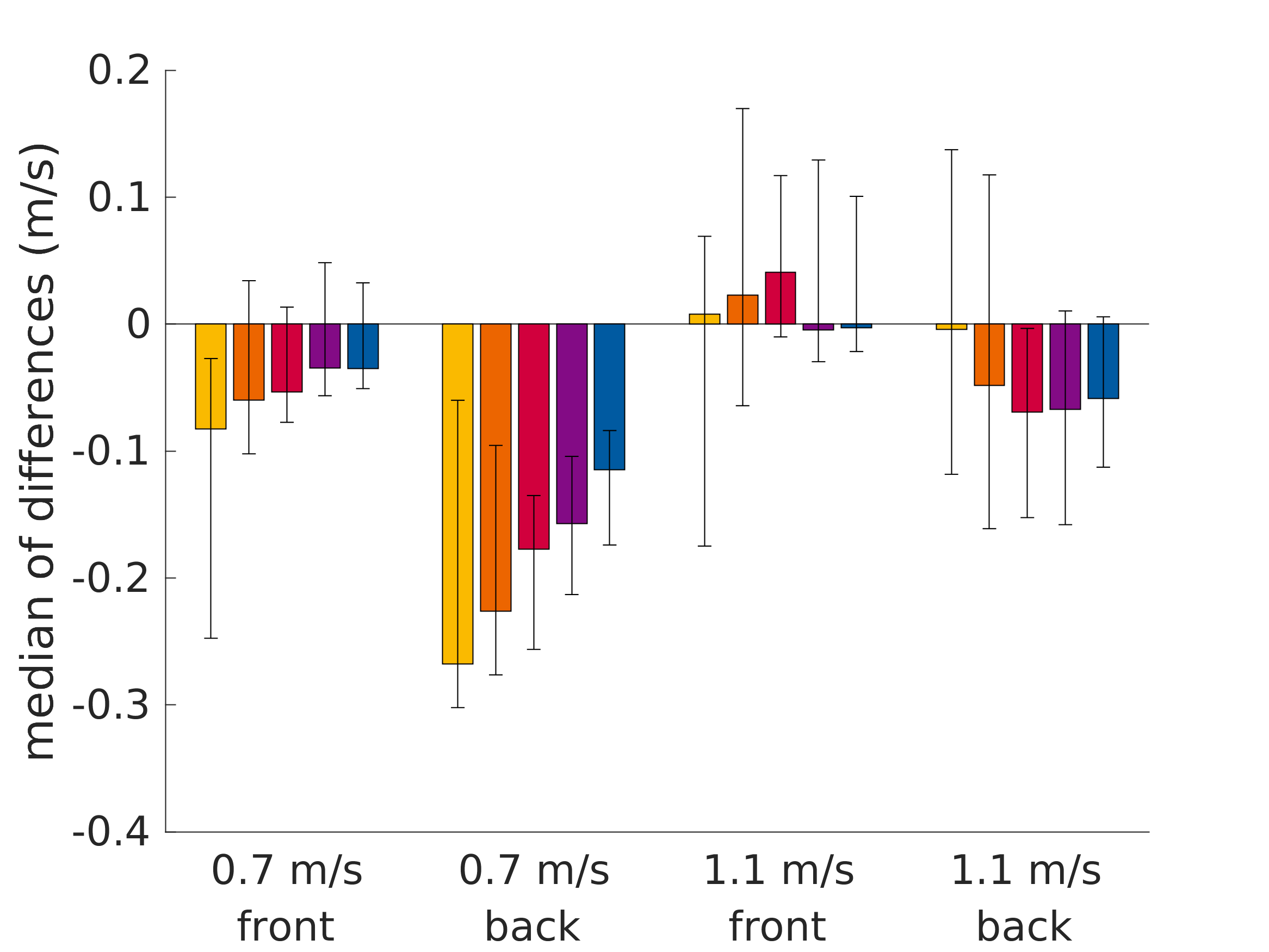}%
			\label{subfig:maxankle}}\\ \vspace{-1em}
  		\subfloat[maximal knee velocity]{\includegraphics[clip,trim= 0 0 20 10,width=0.68\columnwidth]{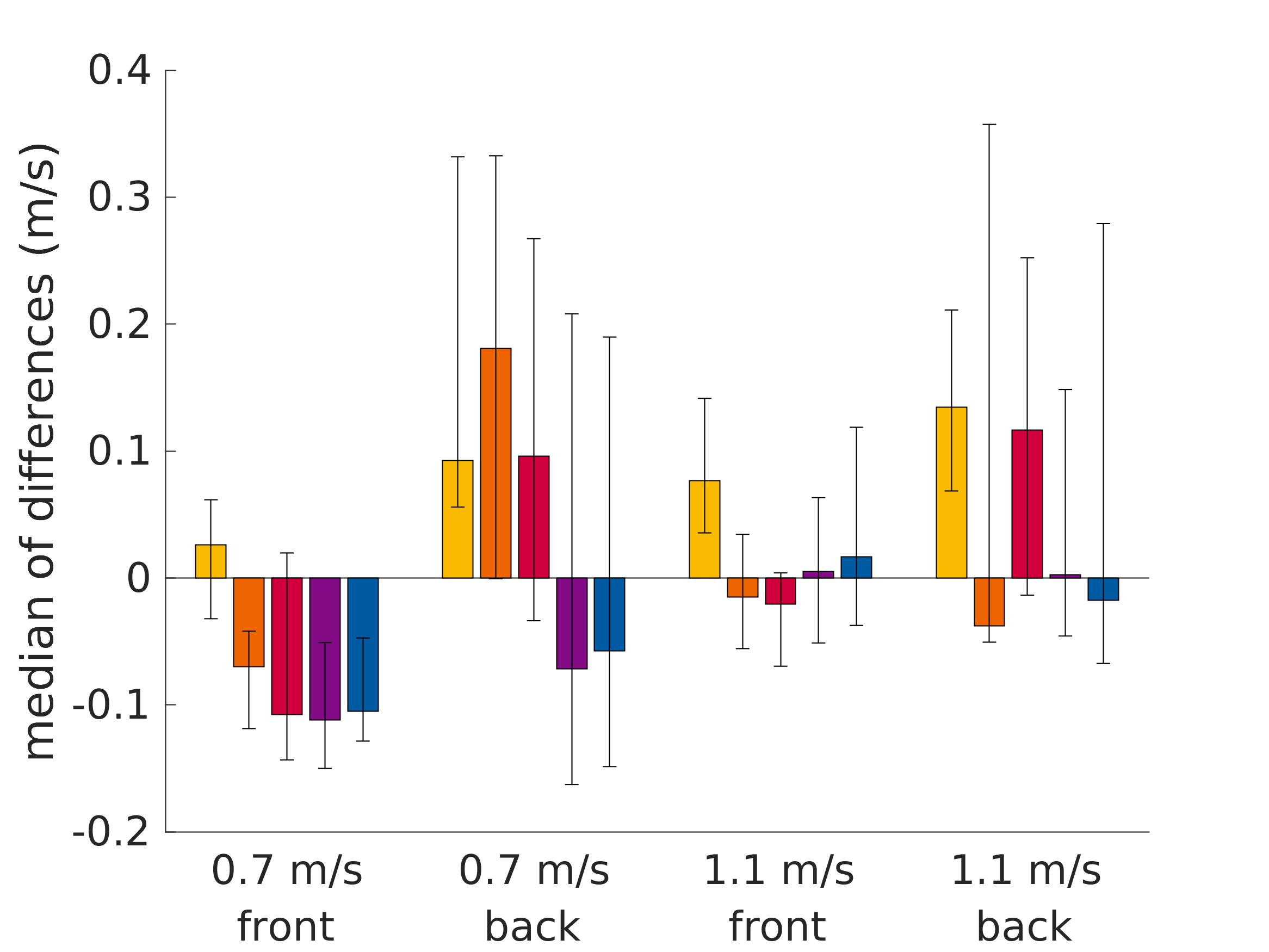}%
			\label{subfig:maxknee}}\hfil
		\subfloat[time instant of maximal knee velocity]{\includegraphics[clip,trim= 0 0 0 10,width=0.68\columnwidth]{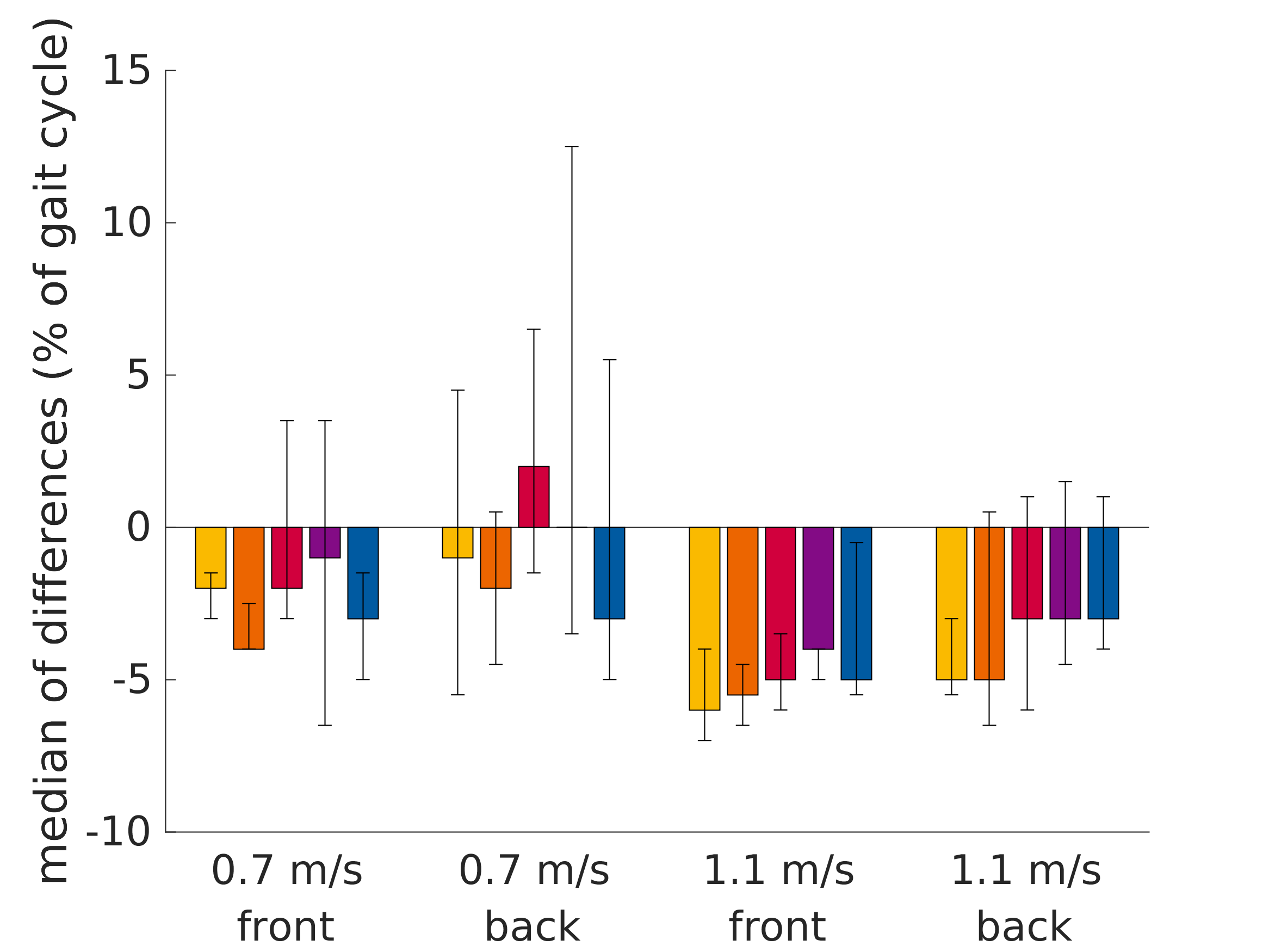}%
			\label{subfig:idxkneemax}}
	}
    \caption{Detailed results for the analyzed gait parameters. The bars indicate the median of differences for the five different knee angle restrictions, the two treadmill speeds and the two radar positions. The error bars indicate the $95$\% confidence intervals of the median. If the confidence intervals include the zero median, we cannot reject the null hypothesis that the median of differences is zero ($p \leq 0.05$).}
    \label{fig:med_gait_paras}
\end{figure*}

\cref{fig:med_gait_paras} shows the estimated median values of the differences $d_i$ and corresponding $95\%$ confidence intervals for the analyzed gait parameters. In order to put the amplitudes of the median of differences into relation, \cref{tab:parameter_avg} shows the corresponding mean values of the gait parameters. \cref{subfig:stridetime} indicates that the stride time can be measured more accurately by the radar for the fast walking speed. Accordingly, the median of differences for the stride length are smaller for $1.1$\,m/s, as shown in \cref{subfig:stridelength}. 
The relative measurement errors for these two gait parameters are very small. From \cref{subfig:steptime}, it can be seen that, similar to the stride time, the step time is more accurately measured by the radar at $1.1$\,m/s. Similarly, the derived step length reveals small measurement errors at both walking speeds, as shown in \cref{subfig:steplength}. The median values of differences for the stance time and the flight time are shown in \cref{subfig:stancetime,subfig:flighttime}, respectively. From \cref{subfig:idxkneemax}, we can see that the front radar overestimates the time index of maximal knee velocity more severely in case of the fast treadmill speed compared to the slow one, which leads to a negative median value of differences. Thus, at $1.1$\,m/s, the stance time is overestimated by radar, while consequently the flight time is underestimated. \cref{subfig:cadence} shows that the cadence can be measured by the radar with high accuracy, since the median of differences assume very small values. As shown in \cref{subfig:maxfoot,subfig:maxankle,subfig:maxknee}, the maximal velocity of the toe, ankle and knee joint reveal absolute median values that are smaller than $0.3$\,m/s. When the radar has a back view on the subject, the maximal knee velocity is measured less reliably, which can be concluded from the large confidence intervals in \cref{subfig:maxknee}. 

\begin{figure}[h!]
  	\centering{
	    \subfloat[$0.7$\,m/s, front radar]{\includegraphics[clip,trim= 0 22 30 15,width=0.69\columnwidth]{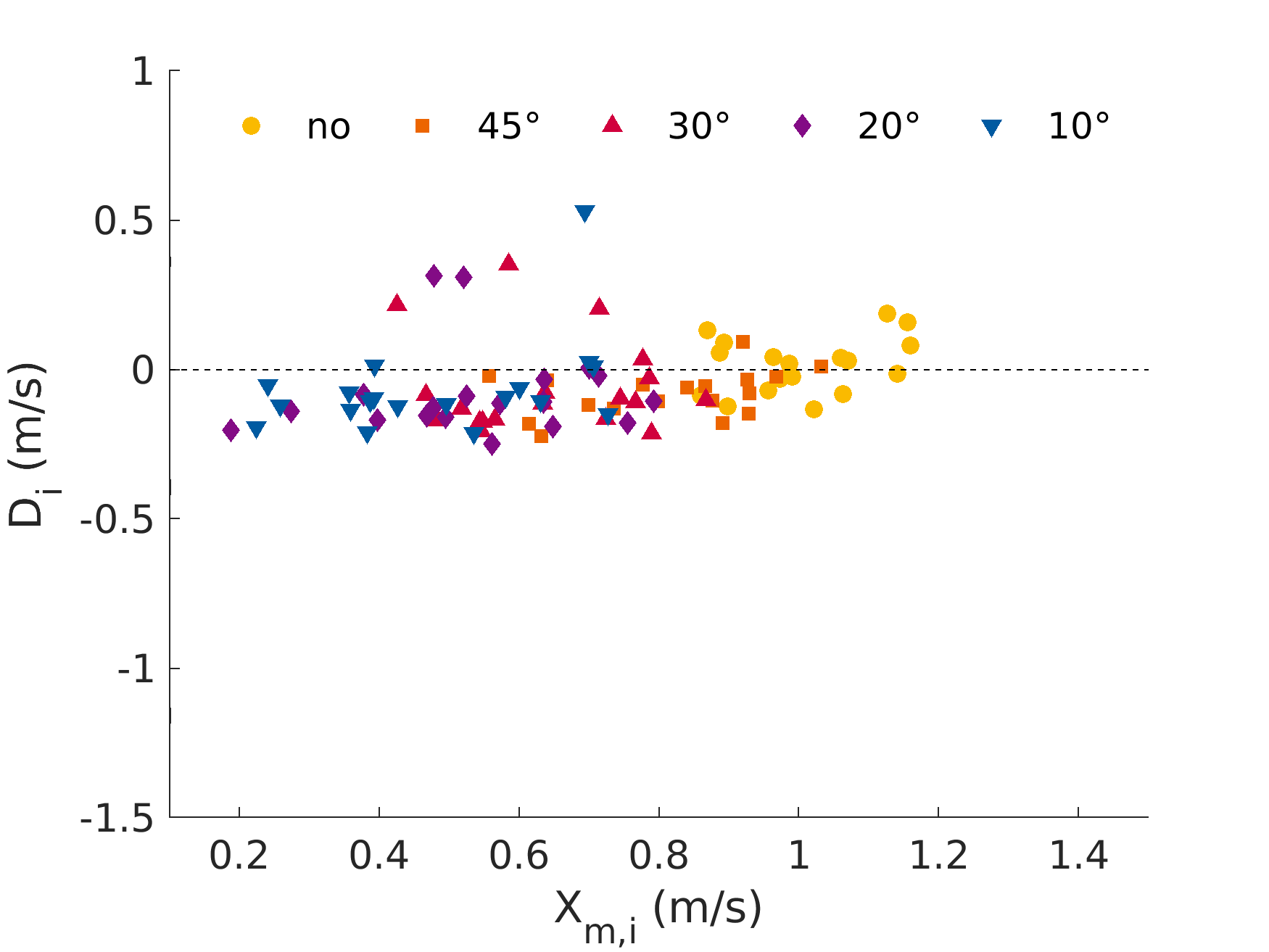}%
			\label{subfig:scat10slowtoward}}\\ \vspace{-0.65em}
		\subfloat[$0.7$\,m/s, back radar]{\includegraphics[clip,trim= 0 22 30 15, width=0.69\columnwidth]{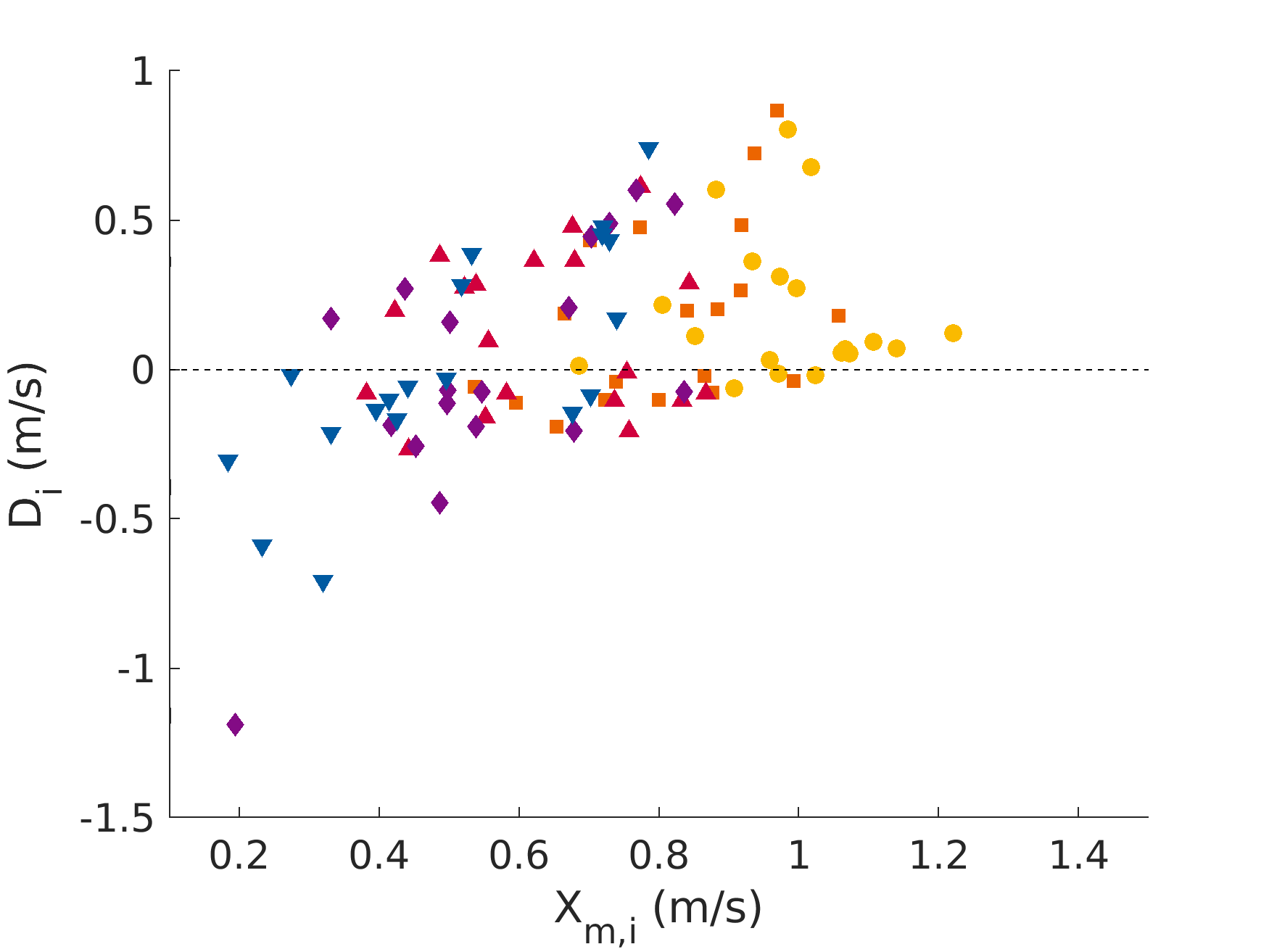}%
			\label{subfig:scat10slowaway}}\\ \vspace{-0.65em}
  		\subfloat[$1.1$\,m/s, front radar]{
  			\includegraphics[clip,trim= 0 22 30 15,width=0.69\columnwidth]{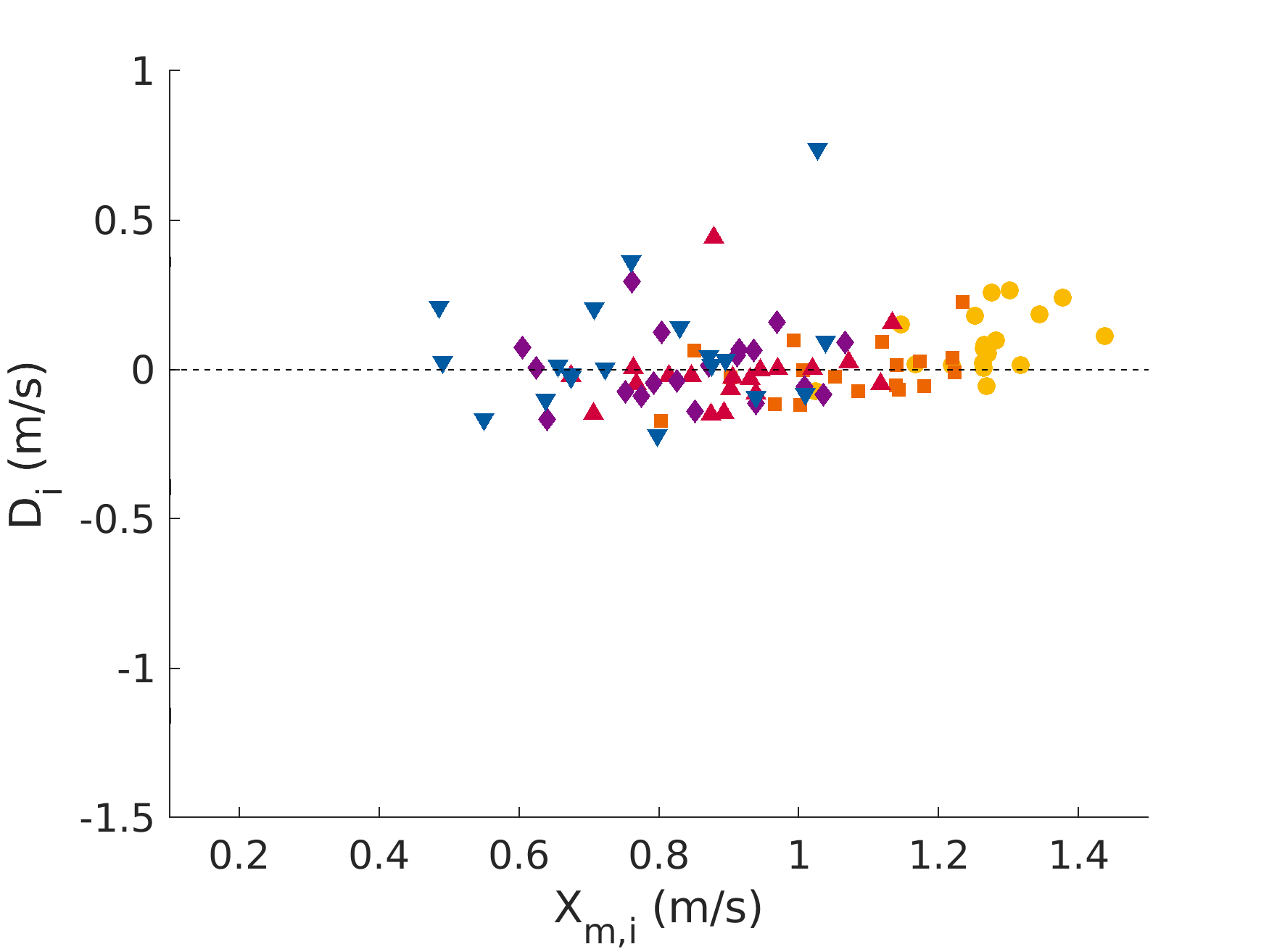}%
			\label{subfig:scat10fastoward}}\\ \vspace{-0.65em}
  	    \subfloat[$1.1$\,m/s, back radar]{\includegraphics[clip,trim= 0 0 30 15,width=0.69\columnwidth]{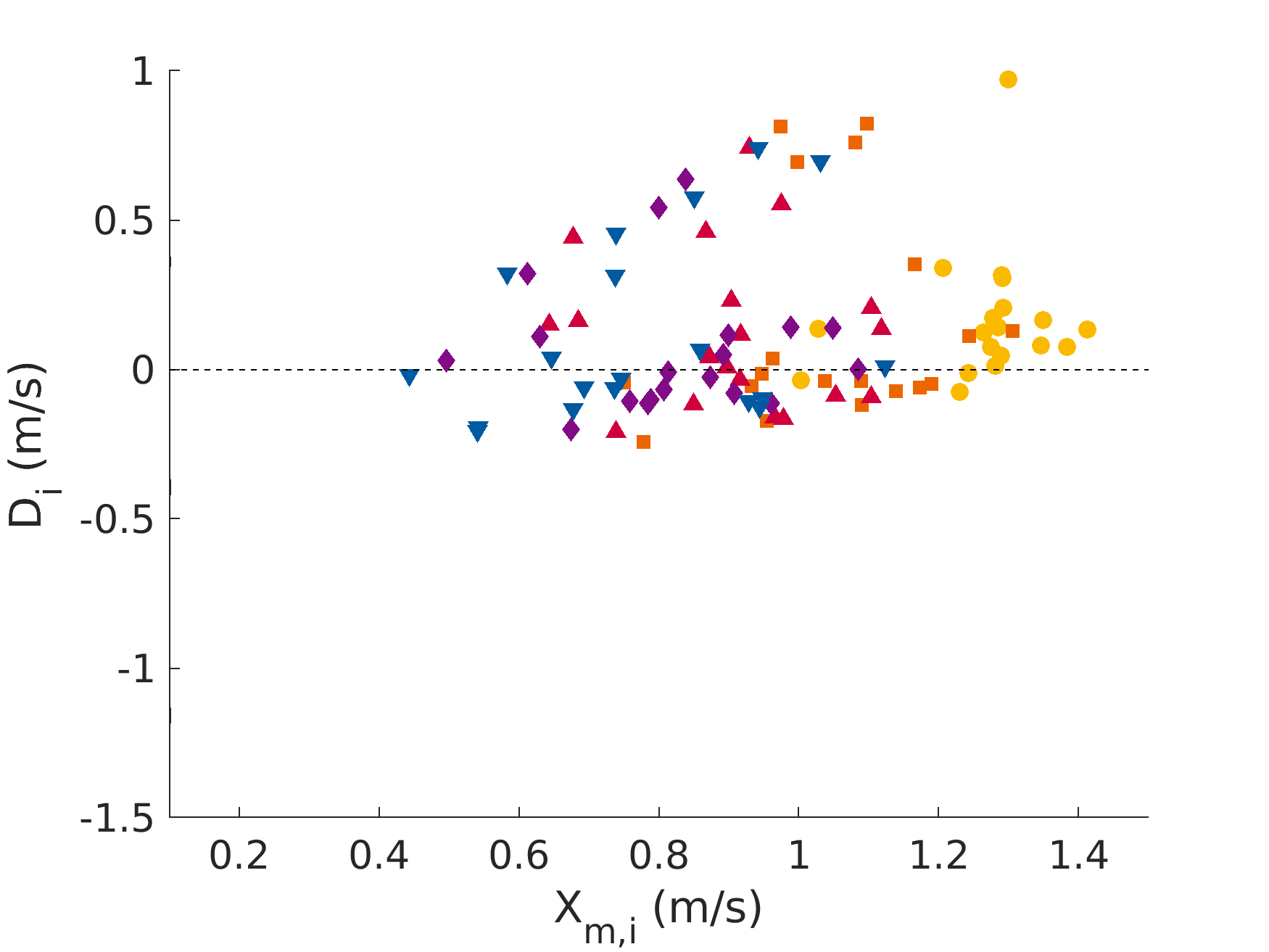}%
			\label{subfig:scat10fastaway}}
	}
    \caption{Decreasing maximal knee velocity for increasing knee angle confinement. The scatter plot shows the differences between motion capture and radar measurements ($D_i$) against the measured values using motion capturing ($X_{m,i}$). The markers represent different degrees of knee angle confinement. The black dashed line indicates that there is no difference between the value measured by the radar and the motion capture system.\label{fig:scatter_knee}}
\end{figure}

Analyzing the last parameter in more detail, \cref{fig:scatter_knee} reveals the decreasing maximal knee velocities for an increasing degree of knee angle restriction (see also parameter 10 in \cref{tab:parameter_avg}). Based on the front radar, the radar measurements are more accurate for all degrees of knee angle confinement with only a few outliers, as shown in \cref{subfig:scat10slowtoward,subfig:scat10fastoward}. \cref{subfig:scat10slowaway,subfig:scat10fastaway} show that the radar measurements are less accurate when observing the subject from behind such that the obtained differences are more dispersed.

\begin{table*}[t!]
\centering
\setlength{\tabcolsep}{2pt} 
\setlength{\lengtha}{3mm}
\caption{Effect of knee angle confinement on the analyzed gait parameters: 1) stride time, 2) stance time, 3) flight time, 4) step time, 5) cadence, 6) stride length, 7) step length, 8) maximal foot velocity, 9) maximal ankle velocity, 10) maximal knee velocity, 11) time instant of maximal knee velocity. For each treadmill speed ($0.7$\,m/s, $1.1$\,m/s) and radar position (front, back), the table shows from left to right the mean values of each parameter for increasing knee restriction (no, $45$\textdegree, $30$\textdegree, $20$\textdegree, $10$\textdegree). The values in the first line are measured by a marker-based motion capture system and values in the second line are obtained using a Doppler radar system.}
\label{tab:parameter_avg}
\resizebox{\textwidth}{!}{%
\begin{tabular}{r *{4}{*{5}{r}@{\extracolsep{\fill}\hskip \lengtha}} c}\toprule
 & \multicolumn{10}{c@{\hskip \lengtha}}{\textbf{0.7\,m/s}} &            \multicolumn{10}{c@{\hskip \lengtha}}{\textbf{1.1\,m/s}} & \multirow{2}{*}{unit}\\
 \cmidrule(l{-0.1em}r{1em}){2-11} \cmidrule(l{-0.1em}r{1em}){12-21}
& \multicolumn{5}{c@{\hskip \lengtha}}{front} & \multicolumn{5}{c@{\hskip \lengtha}}{back}
& \multicolumn{5}{c@{\hskip \lengtha}}{front} & \multicolumn{5}{c@{\hskip \lengtha}}{back}\\\midrule[\heavyrulewidth]
\multirow{2}{*}{1)} & $1.24$ 	& $1.24$ 	& $1.23$ 	& $1.23$ 	& $1.23$ 	& $1.25$ 	& $1.25$ 	& $1.28$ 	& $1.27$ 	& $1.26$ 	& $1.04$ 	& $1.03$ 	& $1.02$ 	& $1.02$ 	& $1.02$ 	& $1.04$ 	& $1.03$ 	& $1.05$ 	& $1.05$ 	& $1.03$ 	& \multirow{2}{*}{s} \\ \medskip 
& $1.24$ 	& $1.24$ 	& $1.23$ 	& $1.23$ 	& $1.23$ 	& $1.25$ 	& $1.25$ 	& $1.28$ 	& $1.27$ 	& $1.26$ 	& $1.04$ 	& $1.03$ 	& $1.02$ 	& $1.02$ 	& $1.02$ 	& $1.04$ 	& $1.03$ 	& $1.05$ 	& $1.05$ 	& $1.03$ 	& \\ 
\multirow{2}{*}{2)} & $0.81$ 	& $0.80$ 	& $0.78$ 	& $0.78$ 	& $0.78$ 	& $0.81$ 	& $0.80$ 	& $0.82$ 	& $0.80$ 	& $0.79$ 	& $0.64$ 	& $0.64$ 	& $0.63$ 	& $0.63$ 	& $0.62$ 	& $0.64$ 	& $0.63$ 	& $0.64$ 	& $0.63$ 	& $0.63$ 	& \multirow{2}{*}{s} \\ \medskip 
& $0.81$ 	& $0.80$ 	& $0.77$ 	& $0.78$ 	& $0.79$ 	& $0.81$ 	& $0.80$ 	& $0.77$ 	& $0.76$ 	& $0.77$ 	& $0.68$ 	& $0.67$ 	& $0.65$ 	& $0.66$ 	& $0.64$ 	& $0.66$ 	& $0.63$ 	& $0.66$ 	& $0.64$ 	& $0.63$ 	& \\ 
\multirow{2}{*}{3)} & $0.42$ 	& $0.44$ 	& $0.44$ 	& $0.45$ 	& $0.45$ 	& $0.44$ 	& $0.45$ 	& $0.46$ 	& $0.46$ 	& $0.47$ 	& $0.39$ 	& $0.40$ 	& $0.40$ 	& $0.39$ 	& $0.40$ 	& $0.40$ 	& $0.40$ 	& $0.41$ 	& $0.42$ 	& $0.41$ 	& \multirow{2}{*}{s} \\ \medskip 
& $0.43$ 	& $0.44$ 	& $0.45$ 	& $0.45$ 	& $0.43$ 	& $0.45$ 	& $0.45$ 	& $0.51$ 	& $0.51$ 	& $0.48$ 	& $0.36$ 	& $0.36$ 	& $0.37$ 	& $0.37$ 	& $0.38$ 	& $0.38$ 	& $0.40$ 	& $0.39$ 	& $0.40$ 	& $0.40$ 	& \\ 
\multirow{2}{*}{4)} & $0.62$ 	& $0.61$ 	& $0.61$ 	& $0.60$ 	& $0.60$ 	& $0.63$ 	& $0.62$ 	& $0.63$ 	& $0.62$ 	& $0.61$ 	& $0.52$ 	& $0.51$ 	& $0.51$ 	& $0.50$ 	& $0.50$ 	& $0.52$ 	& $0.51$ 	& $0.52$ 	& $0.51$ 	& $0.51$ 	& \multirow{2}{*}{s} \\ \medskip 
& $0.62$ 	& $0.61$ 	& $0.61$ 	& $0.61$ 	& $0.61$ 	& $0.62$ 	& $0.62$ 	& $0.64$ 	& $0.63$ 	& $0.63$ 	& $0.52$ 	& $0.52$ 	& $0.51$ 	& $0.50$ 	& $0.50$ 	& $0.52$ 	& $0.51$ 	& $0.52$ 	& $0.52$ 	& $0.51$ 	& \\ 
\multirow{2}{*}{5)} & $97.64$ 	& $97.37$ 	& $98.37$ 	& $98.15$ 	& $98.69$ 	& $96.38$ 	& $96.49$ 	& $94.24$ 	& $95.02$ 	& $95.88$ 	& $116.02$ 	& $116.32$ 	& $117.73$ 	& $117.92$ 	& $118.07$ 	& $116.01$ 	& $116.47$ 	& $114.56$ 	& $114.72$ 	& $116.41$ 	& \multirow{2}{*}{steps/min} \\ \medskip 
& $97.63$ 	& $97.36$ 	& $98.35$ 	& $98.17$ 	& $98.67$ 	& $96.42$ 	& $96.47$ 	& $94.26$ 	& $95.03$ 	& $95.88$ 	& $116.02$ 	& $116.31$ 	& $117.73$ 	& $117.91$ 	& $118.08$ 	& $116.01$ 	& $116.46$ 	& $114.55$ 	& $114.71$ 	& $116.39$ 	& \\ 
\multirow{2}{*}{6)} & $86.52$ 	& $86.57$ 	& $85.82$ 	& $85.99$ 	& $86.05$ 	& $87.64$ 	& $87.52$ 	& $89.56$ 	& $88.71$ 	& $88.09$ 	& $114.20$ 	& $113.81$ 	& $112.56$ 	& $112.32$ 	& $112.27$ 	& $114.13$ 	& $113.65$ 	& $115.59$ 	& $115.29$ 	& $113.74$ 	& \multirow{2}{*}{cm} \\ \medskip 
& $86.53$ 	& $86.58$ 	& $85.79$ 	& $85.99$ 	& $86.05$ 	& $87.60$ 	& $87.54$ 	& $89.53$ 	& $88.71$ 	& $88.10$ 	& $114.21$ 	& $113.82$ 	& $112.56$ 	& $112.33$ 	& $112.25$ 	& $114.14$ 	& $113.65$ 	& $115.59$ 	& $115.31$ 	& $113.74$ 	& \\ 
\multirow{2}{*}{7)} & $43.17$ 	& $42.96$ 	& $42.41$ 	& $42.24$ 	& $42.07$ 	& $43.77$ 	& $43.62$ 	& $44.20$ 	& $43.62$ 	& $42.97$ 	& $56.82$ 	& $56.58$ 	& $55.75$ 	& $55.37$ 	& $54.89$ 	& $56.80$ 	& $56.51$ 	& $57.12$ 	& $56.64$ 	& $55.73$ 	& \multirow{2}{*}{cm} \\ \medskip 
& $43.11$ 	& $43.02$ 	& $42.69$ 	& $42.52$ 	& $42.40$ 	& $43.49$ 	& $43.30$ 	& $44.52$ 	& $43.97$ 	& $43.88$ 	& $57.14$ 	& $56.70$ 	& $55.89$ 	& $55.44$ 	& $55.05$ 	& $56.85$ 	& $56.12$ 	& $57.27$ 	& $56.99$ 	& $56.29$ 	& \\ 
\multirow{2}{*}{8)} & $2.00$ 	& $2.01$ 	& $1.96$ 	& $1.95$ 	& $1.92$ 	& $2.00$ 	& $2.00$ 	& $1.94$ 	& $1.92$ 	& $1.89$ 	& $2.63$ 	& $2.67$ 	& $2.65$ 	& $2.63$ 	& $2.59$ 	& $2.63$ 	& $2.65$ 	& $2.59$ 	& $2.54$ 	& $2.56$ 	& \multirow{2}{*}{m/s} \\ \medskip 
& $2.02$ 	& $2.00$ 	& $1.98$ 	& $1.96$ 	& $1.92$ 	& $2.04$ 	& $2.05$ 	& $2.02$ 	& $1.98$ 	& $1.96$ 	& $2.55$ 	& $2.54$ 	& $2.56$ 	& $2.54$ 	& $2.51$ 	& $2.54$ 	& $2.60$ 	& $2.61$ 	& $2.55$ 	& $2.57$ 	& \\ 
\multirow{2}{*}{9)} & $1.59$ 	& $1.65$ 	& $1.68$ 	& $1.69$ 	& $1.69$ 	& $1.61$ 	& $1.66$ 	& $1.66$ 	& $1.67$ 	& $1.68$ 	& $2.14$ 	& $2.22$ 	& $2.25$ 	& $2.26$ 	& $2.24$ 	& $2.16$ 	& $2.22$ 	& $2.21$ 	& $2.19$ 	& $2.24$ 	& \multirow{2}{*}{m/s} \\ \medskip 
& $1.61$ 	& $1.61$ 	& $1.66$ 	& $1.62$ 	& $1.64$ 	& $1.82$ 	& $1.80$ 	& $1.83$ 	& $1.80$ 	& $1.79$ 	& $2.12$ 	& $2.09$ 	& $2.08$ 	& $2.09$ 	& $2.07$ 	& $2.21$ 	& $2.19$ 	& $2.23$ 	& $2.18$ 	& $2.20$ 	& \\ 
\multirow{2}{*}{10)} & $1.00$ 	& $0.81$ 	& $0.64$ 	& $0.54$ 	& $0.48$ 	& $0.98$ 	& $0.81$ 	& $0.63$ 	& $0.56$ 	& $0.51$ 	& $1.26$ 	& $1.07$ 	& $0.90$ 	& $0.85$ 	& $0.77$ 	& $1.27$ 	& $1.05$ 	& $0.90$ 	& $0.82$ 	& $0.77$ 	& \multirow{2}{*}{m/s} \\ \medskip 
& $0.99$ 	& $0.89$ 	& $0.70$ 	& $0.62$ 	& $0.54$ 	& $0.78$ 	& $0.64$ 	& $0.51$ 	& $0.56$ 	& $0.49$ 	& $1.17$ 	& $1.08$ 	& $0.91$ 	& $0.84$ 	& $0.71$ 	& $1.07$ 	& $0.90$ 	& $0.78$ 	& $0.76$ 	& $0.66$ 	& \\ 
\multirow{2}{*}{11)} & $63.44$ 	& $61.28$ 	& $62.47$ 	& $61.68$ 	& $61.47$ 	& $63.53$ 	& $61.58$ 	& $62.84$ 	& $63.37$ 	& $62.16$ 	& $60.11$ 	& $59.22$ 	& $59.79$ 	& $59.79$ 	& $59.21$ 	& $59.95$ 	& $58.47$ 	& $60.21$ 	& $59.74$ 	& $59.11$ 	& \multirow{2}{*}{\%} \\ \medskip 
& $65.50$ 	& $64.56$ 	& $63.16$ 	& $63.00$ 	& $64.63$ 	& $64.47$ 	& $63.74$ 	& $60.21$ 	& $59.74$ 	& $61.58$ 	& $65.61$ 	& $64.72$ 	& $63.89$ 	& $64.26$ 	& $62.26$ 	& $63.32$ 	& $61.58$ 	& $62.42$ 	& $61.37$ 	& $60.95$ 	& \\ 
\bottomrule
\end{tabular}}
\end{table*}

\subsection{Discussion}
Using motion capture data, we validated existing and new methods for extracting a variety of medically relevant gait parameters from experimental radar data. 

Utilizing the standard micro-Doppler envelope, five spatiotemporal parameters related to rhythm (stride time, step time, cadence) and pace (stride length, step length) are extracted from the radar measurements. For these parameters, the differences between the values obtained through radar and motion capturing are close to zero for all analyzed conditions. 

Additional envelopes are extracted from the micro-Doppler signatures to measure kinematic parameters, i.e., the maximal velocity of the toe, ankle and knee joint during walking. For different knee angle confinements, the velocities measured by the radar system agree with those obtained using motion capturing in $51$ out of the $60$ analyzed cases ($85$\%). In particular, the changes in maximal knee velocity due to the restriction of the knee's motions are also captured by the radar. Here, the thresholds for extracting the envelope signals were found empirically. In general, the appearance of the micro-Doppler signatures depends on the subject's stature (height, mass), interfering noise component due to e.g.~multi-path, and transmitting power of the radar system, which can result in different thresholds. Here, we note that the presented micro-Doppler signatures might be affected by the backscatterings from the orthosis. However, preliminary tests showed that the differences to not wearing an orthosis are negligible. 

The knee's micro-Doppler envelope is further utilized to detect the time instant of maximal knee velocity. Due to the proposed definition of the flight time (see \cref{sec:flight_time_definition}), the radar-based stance and flight time measurements rely on accurate detection of the time instant at which the knee reaches its maximal velocity during the gait cycle. Since radar micro-Doppler signatures are composed of overlaying reflections from different body parts, the knee's signatures is obscured by backscatterings from the upper leg (mid upper leg to knee joint), which exhibits similar velocities as the knee joint while reflecting more energy of the \ac{EM} wave. From \cref{rad_normal_fast_front,rad_abnormal_fast_front,rad_normal_slow_front}, it can be seen that in case the radar has a front view on the target, the toe and ankle signatures can be identified by the red areas that exhibit a sinusoidal shape between $70$\% and $100$\% of the gait cycle. When compared to \cref{moc_normal_fast_front,moc_abnormal_fast_front,moc_normal_slow_front}, the measured radial velocities generally align well with the recorded velocities by the motion capture system. However, when that radar is positioned behind the subject, the knee's signature does not exhibit the shape as proposed by the motion capture data, but appears to be much more complex, as shown in \cref{rad_normal_fast_back}. Here, an almost vertical line, also referred to as spike signature, appears at approximately $85$\% of the gait cycle in \cref{rad_normal_fast_back}. These differences have first been reported in \cite{Sei17} and are due to the different backscattering characteristics of the lower leg from the front as compared to from the back. Thus, the precise measurement of the knee joint velocity during walking remains challenging using micro-Doppler signatures. In order to extract kinematic parameters more reliably from radar data, averaged micro-Doppler signatures, such as the ones shown in \cref{rad_normal_fast_front,rad_abnormal_fast_front,rad_normal_slow_front}, should be used as they are less corrupted by noise and hence more conclusive. Further, model-based approaches could be considered that are specifically designed to extract kinematic parameters by incorporating prior knowledge. 

The presented results indicate the feasibility of using Doppler radar for unobtrusively measuring a variety of biomechanical parameters that can be of use for basic gait analysis. The proposed methods need to be further evaluated using realistic in-home data without a treadmill and a larger clinical population. In general, treadmill walking differs from walking overground \cite{Hol16}. However, in terms of the here investigated gait parameters, i.e., spatiotemporal parameters and lower limb kinematics in the sagittal plane, studies suggest that there are only few or no differences (in mean values) between treadmill and overground walking (see e.g.~\cite{Hol16,Lee08}). In this regard, we note that most \textit{spatiotemporal} parameters can also be extracted from realistic radar data from overground walking, as e.g.~demonstrated by Wang \textit{et al.} \cite{Wan14}. However, the extraction of \textit{kinematic} parameters is more sensitive to the distance of the person to the radar and the viewing angle of the radar (vertical/horizontal) on the subject, who might not walk along the radar's \ac{LOS}. Thus, while this work focused on the lower limbs' motions only, future studies should reevaluate the positioning of the radar, i.e., its distance to the subject and height, which directly affects the appearance of the micro-Doppler signatures. Further studies may also consider gait abnormalities that impact a larger number of gait parameters (more drastically). For example, opposed to only restricting the knee angle during walking, the ankle's motion could be constrained, too. This could emulate trans-tibia and trans-femoral amputees, respectively, where the latter induces a change in gait parameters compared to a healthy leg \cite{Nol03}.  

Finally, though the measured \textit{absolute} values obtained through radar measurements ought not be sufficiently accurate for professional gait analysis (yet), we showed that radar is capable of sensing small changes in gait parameters induced by the orthosis, which can be of interest for assessing the \textit{relative} progress of diseases affecting the gait\cite{Gho17}. Additionally, asymmetry measures could be calculated and used to detect and quantify gait impairments, as suggested in e.g.~\cite{Gou17,Nol03}. Gait asymmetry measures typically depend on the difference in gait parameters from the left and right leg, rather than on their absolute values. Based on experimental radar measurements of $6$\,s duration, the gait asymmetry of four gait-impaired individuals was successfully detected in \cite{Sei19a}. A sequential detection approach for increasing recognition speed and thus practicability, was presented in \cite{Sei19b}. Further, deep learning methods could be utilized to automatically learn the relation between biomechanical parameters and radar micro-Doppler stride representations, without relying on specifically designed feature extraction methods. For example, Hannik \textit{et al.} \cite{Han17} proposed such a framework to extract eight spatiotemporal stride parameters from inertial sensors using deep convolutional neural networks. Hence, a radar-based long-term gait analysis system could provide useful information, e.g.~on the progress of recovery from an injury in rehabilitation, to assist medical practitioners and physiotherapist in therapy. 

\section{Conclusion}\label{sec:conclusion}
The study investigated if Doppler radar is capable of extracting biomechanical gait parameters in a similar quality as marker-based motion capturing. New methods were introduced to measure the flight time and lower limb joint linear velocities based on radar micro-Doppler signatures. Existing and new methods for extracting spatiotemporal and kinematic parameters were evaluated in unimpaired gait and for gait abnormalities introduced by an orthosis, at different walking speeds, and at a frontal and a rear positioning of the radar. Based on experimental data of nineteen volunteers walking on a treadmill, the gait parameters obtained through radar and motion capturing were qualitatively compared and differences were statistically analyzed. Five spatiotemporal and three kinematic parameters could be extracted from the radar data with high accuracy for most of the considered conditions. However, in the proposed method, flight and stance time measurements rely on the correct detection of the time instant of maximal knee velocity. In conclusion, radar presents a promising technology for unobtrusive in-home gait analysis, which could supplement existing tools and thus, e.g., contribute to aid healthcare professionals in diagnosis and in the course of treatment.

\bibliographystyle{IEEEtran} 
\bibliography{ms}

\end{document}